# Polymorphism of two-dimensional antiferromagnets, $AgF_2$ and $CuF_2$

**Daniel Jezierski* and Wojciech Grochala***

*d.jezierski@cent.uw.edu.pl, w.grochala@cent.uw.edu.pl

*Center of New Technologies, University of Warsaw, 02089 Warsaw Poland*

This work is dedicated to Professor Jacek Klinowski at his $80^{th}$ birthday

## Abstract

We present theoretical study of relative stability as well as of the magnetic and electronic properties of $AgF_2$ and $CuF_2$, in two related structural forms: orthorhombic (ambient pressure form of $AgF_2$) and monoclinic (ambient pressure form of $CuF_2$), using Density Functional Theory. We show, that at $P2_1/c$ → *Pbca* structural transition is associated with weakening of intra-sheet magnetic superexchange. This finding is consistent with the flattening of 2D layers, smaller charge-transfer energy and stronger admixing of $Ag_d/Cu_d$-$F_p$ states in monoclinic structure, comparing to orthorhombic form. Consequently, monoclinic $AgF_2$ should be targeted in experiment as it should show stronger magnetic coupling than its orthorhombic sibling. The dynamically stable $P2_1/c$ form of $AgF_2$ could be achieved via two alternative paths: by applied negative strain, or by rapid quenching silver(II) difluoride from temperatures higher than 480 K to low temperatures.

## Graphical abstract

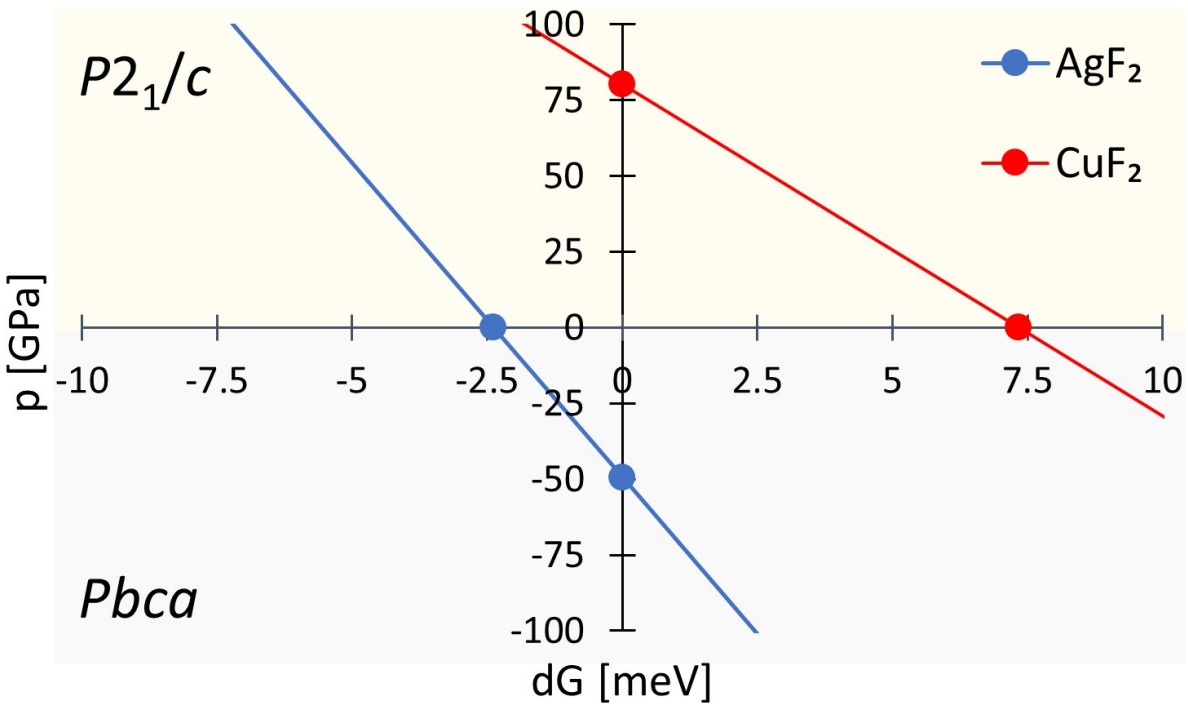

## Keywords

layered materials, antiferromagnets, polymorphism, pressure, phase diagram

## Introduction

The most typical structure adopted by transition metal difluorides at ambient pressure is rutile type (Fig. 1). For $TMF_2$, where TM = V, Mn, Fe, Co, Ni, Pd, Zn, difluorides crystallize in tetragonal system $P4_2/mnm$ space group[1–3]. Building blocks in this structure consist of nearly regular octahedra of fluorine atoms around each metal ion (CN=6), with both edge and corner sharing of F atoms. However, in the case of $CrF_2$[4] and $CuF_2$[5], the deformation of octahedra is pronounced, predominantly due to the Jahn-Teller effect. Here, the elongation of apical M-F bonds leads to the symmetry lowering, with the resulting monoclinic structure ($P2_1/c$). Another important structural type among metal difluorides is fluorite with a cubic ligand environment around metal cation (CN=8). However, among TM systems, this type is adopted only by three difluorides: two with closed-shell cations - $CdF_2$, $HgF_2$ (which could be classified as post-transition ones) and only one with a genuine open d-shell cation – $AgF_2$. For the latter, Jahn-Teller effect drives the distortion from the parent cubic polytype towards the orthorhombic one (*Pbca*)[6]. Thus, both $CuF_2$ and $AgF_2$ exemplify the unique low-symmetry structures types, otherwise unknown among ionic fluorides.

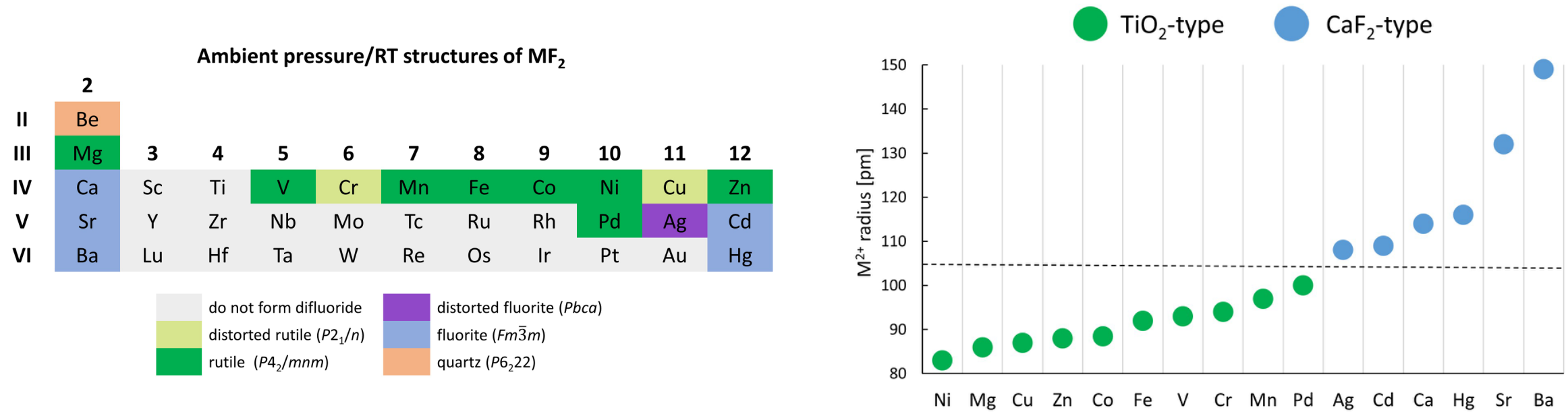

**Fig. 1** Ambient pressure/room temperature structure of selected $MF_2$ (left) and type of structure adopted of binary difluorides depending on the $M^{2+}$ ionic radius in octahedral environment (right). The dashed line indicates the boundary between two structure types: rutile and fluorite.

Polymorphism of metal difluorides is even richer at elevated pressure conditions, as extensively studied during the last years[3,7–14]. Most TM difluorides adopt close-packed cotunnite (*Pnma*) structure (CN=9) upon compression, with $Ni_2In$-type (CN=11) as un ultimate high-pressure structure. For example, $FeF_2$ undergoes the following sequence of phase transitions: rutile → $CaCl_2$ → $Ni_2In$, thus the CN of metal atom increasing from 6 via 9 to 11. Another example is $NiF_2$, where compression up to 3 GPa induces rutile → $CaCl_2$ transformation[10]. However, theoretical study predicts that at 30 GPa this compound should adopt monoclinic structure ($C2/c$), and this remains to be verified in experiment[15]. Additionally, there are two examples of difluorides where cotunnite structure has not yet been observed experimentally. The first example is $PdF_2$ (with high-spin $d^8$ cations) where rutile → cubic ($Pa\overline{3}$ ) transformation occurs at 1.5 GPa[3]; the cubic polytype can be also stabilized in the low-temperature regime[16]. The second one is $AgF_2$ which undergoes the following transitions: P*bca* → P*ca*2$_1$ → P*bcn* → P*nma*[13], with two unusual intermediate forms: a non-centrosymmetric P*ca*2$_1$ one and a unique nanotube-like P*bcn*. A similar sequence is theoretically predicted for its lighter congener, $CuF_2$[11]. Indeed, the $P2_1/c$ → *Pbca* ($AgF_2$-type) transition was observed for $CuF_2$ at 9 GPa, while the $PbCl_2$-type structure is predicted theoretically to occur only at 72 GPa[11]. Clearly, $CuF_2$ and $AgF_2$ behave quite differently from the rest of the TM difluoride herd.

Polymorphism strongly affects key material properties such as e.g. magnetic and electronic ones. We note that both $AgF_2$ and $CuF_2$ are layered, with corrugated sheets constructed from $[MF_4]$ units (Fig.2). Still, the disparate stacking of sheets and quite different nature of M-F bonds in both compounds (more covalent for Ag, more ionic for Cu), strongly affect their physicochemical properties. Silver difluoride in many ways mimics undoped oxocuprates (HTSC precursors)[17–19], and diverse ways of its doping have been theoretically[20–24] and experimentally[25] explored. As a result of these efforts, we have recently observed the coexistence of two structural types of solid solutions in Cu-Ag-F phase diagram: monoclinic ($CuF_2$-type) and orthorhombic (parent $AgF_2$-type) (Fig.2)[25]. Their formation is unexpected due to substantial difference in the cation size ($R_{Cu^{2+}}$ = 0.87 Å and $R_{Ag^{2+}}$ = 1.08 Å) across the rutile/fluorite boundary (Fig.1). Interestingly, despite that, at experiment conditions discussed in Ref.25, as much as 44% of Ag could be doped into $CuF_2$ structure, and up to 30 % Cu into the $AgF_2$ type. The formation of mixed-cation phases raises a natural question: could Cu-free $AgF_2$ be obtained in the monoclinic form, or Ag-free $CuF_2$ in the orthorhombic form, and what would be the key properties of these polymorphs? The experimental results suggest that $CuF_2$ may indeed be pushed into $AgF_2$ type at elevated pressure[9] but it is not known yet whether the high-pressure type is quenchable to ambient (p,T) conditions, and what its properties would be. On the other hand, whether $AgF_2$ could be obtained in the $CuF_2$ type is an open question.

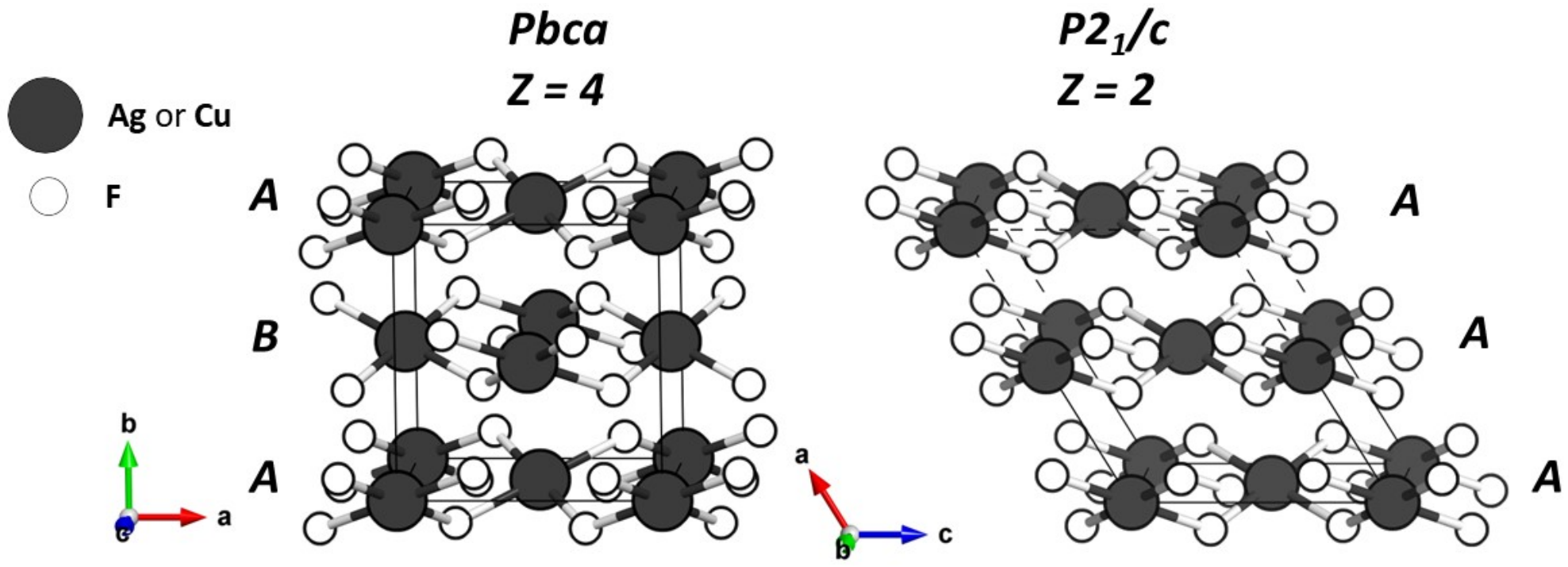

**Fig. 2** The ambient pressure structures of $AgF_2$ *Pbca* (left) and $CuF_2$ *P*$2_1$*/c* (right). Only intra-layer M-F bonds are shown for clarity. *ABA* and *AAA* marks refer to the type of layer stacking. In case of *P*$2_1$*/c* structure, 1x2x1 supercell is presented for direct comparison with the orthorhombic structure.

Motivated by these findings, in this paper we theoretically study the dynamic and thermodynamic stability of four phases i.e. $CuF_2$ and $AgF_2$ each in two different structure types, and we evaluate their key electronic and magnetic properties. Finally, we propose possible experimental pathway to obtain the Cu-free monoclinic $AgF_2$.

**Methods**

Calculations were performed within the DFT approach as implemented in VASP 5.4.4 code[26], using a GGA-type PBEsol[27] functional with projected augmented wave method[28,29] .The on-site Coulombic interactions of d electrons were realized by introducing Hubbard ($U_d$) and Hund ($J_H$) parameters (DFT+U formalism), as proposed by Liechtenstein[30]. In DFT+U and DFT+U+vdW methods, the $J_H$ was set to 1 eV[31], while the Ud value was set to 8 and 10 eV, respectively for Ag and Cu[32,33]; where van der Waals dispersion energy-correction term was provided following the expression introduced by Grimme et al. (DFT-D3)[34]. Two other methods have been also employed: SCAN and HSE06. Plane-wave cutoff energy of 520 eV was used in all systems. The k-spacing was set to 0.032 $Å^{-1}$ for geometry optimization and 0.022 $Å^{-1}$ for self-consistent-field convergence. The convergence thresholds of $10^{-7}$ eV for electronic and $10^{-5}$ ionic steps were used. Calculations of phonon curves for each structure were realized using PHONOPY package[35] (step-by-step procedure is available online, on the official website). ZPE energies/Γ-point frequencies, were calculated using DFPT or finite differences approaches provided by VASP software.

**Results and discussion**

*1.1. Structural features of silver and copper difluorides in both polymorphic forms*

For $AgF_2$ in its *P*bca type, the first coordination sphere of cation, can be described as deformed rhombic prism (2+2+2+2 coordination), while for $CuF_2$ in its parent *P*$2_1$/c structure, as an elongated octahedron (4+2 coordination) (**Fig.3**).

Our theoretical calculations nicely reproduce experimental structural parameters of $AgF_2$ and $CuF_2$ in their most stable structures (**Table 1**). The best agreement between experimental and theoretical volumes of the unit cell was found using SCAN method. The fully optimized $AgF_2$ and $CuF_2$ unit cells differs in volume only by 1.75% and 1.66%, respectively, to experimental ambient pressure and ambient temperature forms[6,36]. Comparable results come from HSE06 method, where volume differ by 1.67% and 1.96%, respectively, for silver and copper difluoride from experimental data. The DFT+U method with van der Waals correction, gives worse agreement – volumes of optimized unit cells are underestimated by 5.51% and 7.02% compared to experimental ones, for $AgF_2$ and $CuF_2$, respectively. However, since all calculations refer to T → 0 K and p → atm 0 conditions, the discrepancies are expected; some are also related to the limitations of these theoretical methods. We note that adjustment of the U and J values, used in the spin polarized DFT calculations, may lead to better compatibility between experimental and theoretical structural parameters (see Tokar et. al[37] and Kurzydłowski[11], where in DFT+U calculations U = 5 and J = 1 for $AgF_2$ and U =7 and J = 0.9 eV for $CuF_2$, respectively, give satisfactory agreement between experimental and theoretical structural parameters).

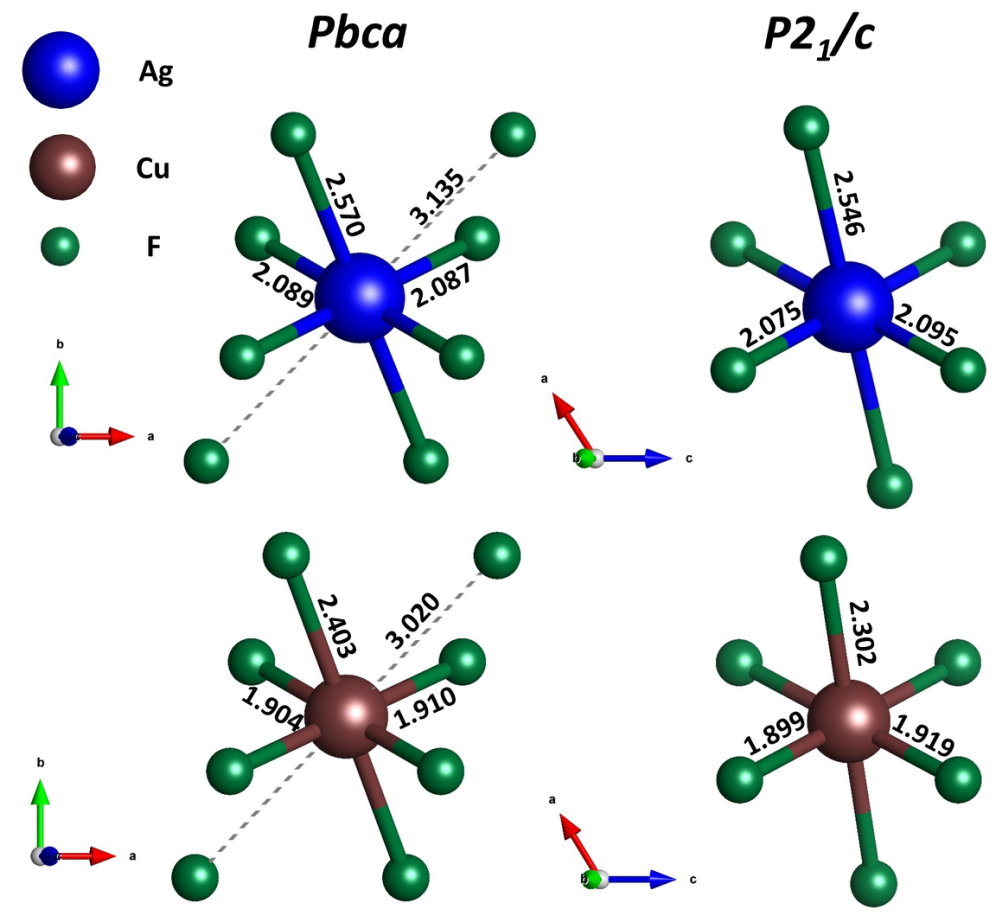

**Fig. 3** First coordination sphere for $AgF_2$ (up) and $CuF_2$ (down) in *Pbca* (left) and *P2₁/c* (right) cells. The M-F bonds lengths are expressed in Å, following SCAN calculations.

**Table 1**. Unit cell parameters *a*, *b*, *c*, volumes V as estimated from experimental measurements (X-ray and neutron powder diffraction, XRPD and NPD) with theoretical values from our calculations for orthorhombic $AgF_2$ and monoclinic $CuF_2$ (for antiferromagnetic spin ground state). For orthorhombic cell all unit cell angles are equal 90˚, where for monoclinic one α = γ = 90˚ ≠ β.

| Method | $AgF_2$ (*Pbca*) | | | | Method | $CuF_2$ ($P2_1/c$) | | | | |
|---|---|---|---|---|---|---|---|---|---|---|
| | *a* [Å] | *b* [Å] | *c* [Å] | V/FU [Å$^3$] | | *a* [Å] | *b* [Å] | *c* [Å] | β [˚] | V/FU [Å$^3$] |
| *PXRD, RT*[25] | 5.550 | 5.836 | 5.095 | 41.257 | *PXRD, RT*[25] | 3.294 | 4.559 | 5.345 | 121.11 | 34.704 |
| *NPD, 4.2K*[6] | 5.524 | 5.780 | 5.066 | 40.438 | *NPD, 4.2K*[36] | 3.294 | 4.568 | 5.358 | 121.17 | 34.491 |
| *XRD, RT*[26] | 5.568 | 5.831 | 5.101 | 41.403 | PXRD, RT[11] | 3.302 | 4.560 | 5.352 | 121.11 | 34.497 |
| *DFT+U* | 5.433 | 5.768 | 5.008 | 39.234 | *DFT+U* | 3.192 | 4.505 | 5.293 | 121.32 | 32.510 |
| *DFT+U+vdW* | 5.408 | 5.660 | 4.993 | 38.211 | *DFT+U+vdW* | 3.185 | 4.479 | 5.286 | 121.72 | 32.068 |
| *SCAN* | 5.583 | 5.724 | 5.151 | 41.145 | *SCAN* | 3.302 | 4.522 | 5.310 | 121.20 | 33.917 |
| *HSE06* | 5.539 | 5.843 | 5.082 | 41.115 | *HSE06* | 3.296 | 4.516 | 5.300 | 120.98 | 33.817 |

Now, we will discuss the key structural differences between *Pbca* and $P2_1/c$ structure type. There are <u>three</u> main structural differences between these structures: **1)** different stacking of $[MF_4]$ plaquettes i.e. ABA *vs.* AAA (**Fig. 1**); **2)** reduction of coordination number for $M^{2+}$ from 8 to 6, and **3)** accompanying changes in the bond lengths and bond angles. Based on SCAN calculations, four intralayer (equatorial) Ag-F bond lengths in $AgF_2$ have similar values: 2x2.087 Å and 2x2.089 Å. The apical (axial) bonds are considerably longer: 2x2.570 Å. The last Ag-F contacts, completing the $Ag^{2+}$ first coordination sphere, are significantly longer than the others: 2x3.135 Å. Similarly, for $CuF_2$ in *Pbca* structure the first coordination sphere of cation includes the shortest contacts of 2x1.904 Å and 2x1.910 Å, the apical ones of 2x2.403 Å, and secondary interactions at 2x3.020 Å.

In the $P2_1/c$ system, the coordination sphere of $M^{2+}$ is 6-fold (CN=6, **Fig. 2**). In $AgF_2$, two intralayer Ag-F contacts are slightly shorter and two slightly longer than in the *P*bca structure: 2x2.075 Å and 2x2.095 Å. Apical bond lengths differ to a larger extent (2x2.546 Å in $P2_1/c$ form vs. 2x2.570 Å+2x3.135 Å) i.e. they are shorter in monoclinic form (which is consistent with the reduction of the CN of the cation). In $CuF_2$ case, the intra-layer copper-fluoride bonds for the $P2_1/c$ form are 1.899 Å and 1.919 Å; however, the apical contacts are substantially shorter for this form (2x2.302 Å *vs.* 2x2.403 Å

+ 2x3.020 Å); this is consistent with the results previously reported by Kurzydłowski[11]. Results from all computational methods indicate similar tendencies of bond length changes despite slight differences in absolute values (Table ESI 1-2).

Puckering of the $MF_2$ layers is also different in both structure types. For example, for the monoclinic $AgF_2$, the intra-layer Ag-F-Ag angle is larger by ca. 1.2˚ than in orthorhombic one (SCAN). For $CuF_2$ the effect is even more pronounced (3.0˚). This may be related to the different metal-metal distances in both structures. In the $P2_1/c$ system Ag-Ag intra-layer distance is larger by 0.3%, whereas inter-layer contact is shorter by 2.8% compared to $AgF_2$ *Pbca* form. For copper difluoride, the intralayer Cu-Cu distances is by 1.3% larger, where interlayer one is by 8.0% shorter for $P2_1/c$ form (all values are listed in Table. SI1-2 in ESI). All these features impact, as we will see, the magnetic and electronic properties of both polymorphs.

## *1.2. Dynamic stability*

To evaluate dynamic stability, we calculated phonon modes at Gamma point for $AgF_2$ and $CuF_2$ in both polymorphic forms: *Pbca* and $P2_1/c$ (Table x ESI). Experimental (IR, RAMAN) and theoretical phonon frequencies of $AgF_2$ (*P*bca) with their symmetry and optical activity, have been previously reported by Gawraczyński, et al[17]. Additionally, Kurzydłowski[11] previously assigned bands RAMAN active bands of $CuF_2$ ($P2_1$/c). Here, we extended this assignment to IR-active $CuF_2$ bands of the monoclinic form. Additionally, we calculated dispersion of phonons for binary difluorides at $P2_1/c$ and *Pbca* structures in DFT+U framework (**Fig. 4**).

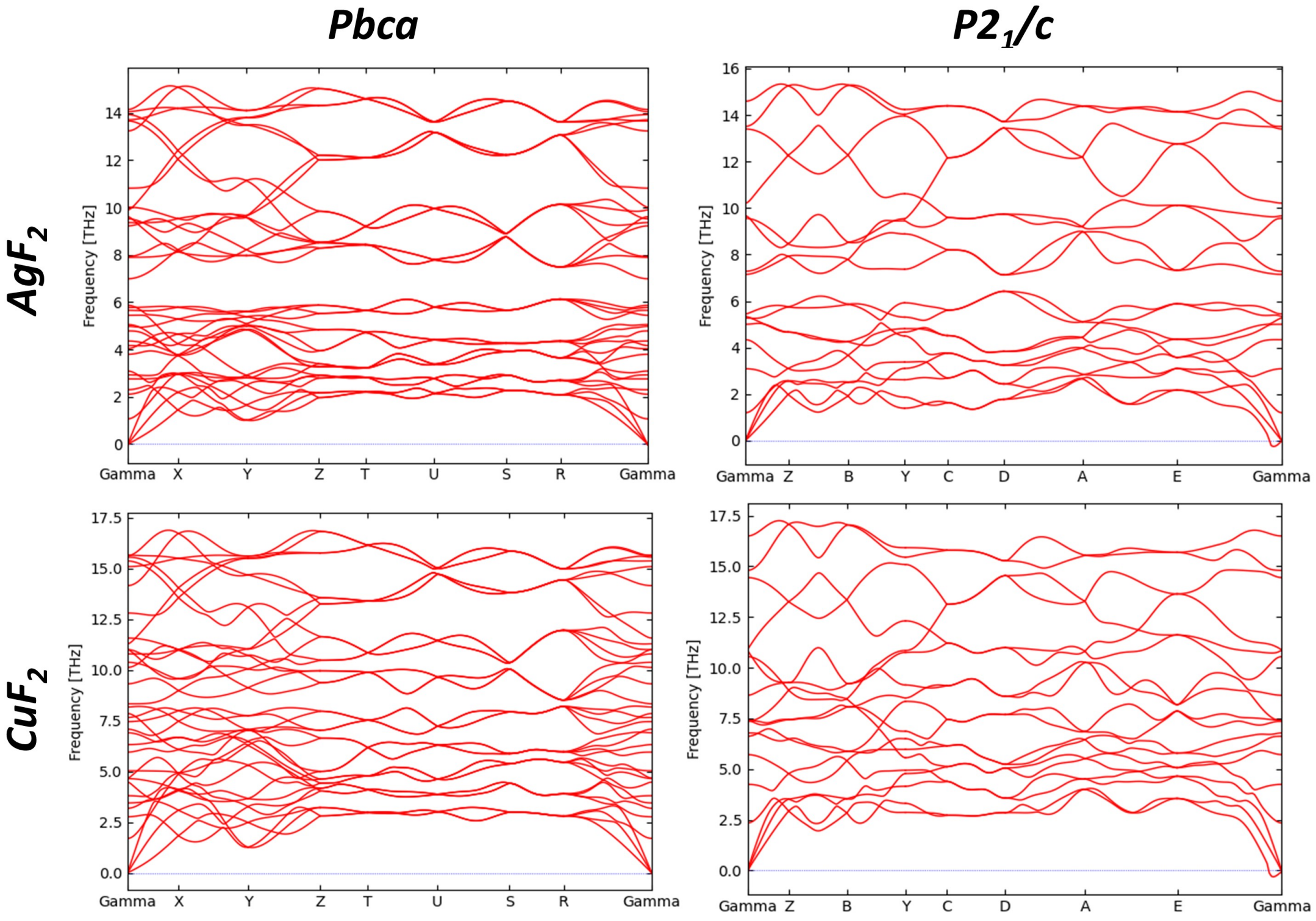

**Figure 4.** Phonon dispersion curves for $AgF_2$ (upper) and $CuF_2$ (lower) in Pbca (left) and $P2_1$/c systems (right).

Analysis of phonon dispersion curves (**Fig.4)** shows no dynamic instability for both compounds in orthorhombic system (P*bca*). However, in monoclinic form ($P2_1/c$), both compounds show very small soft features at q = (-0.04, 0.04, -0.04). This obviously is not an exact in reciprocal space, and indeed, geometry optimization of the structures obtained following the said imaginary mode do not lead to lower-energy (see SI4). In other words, tiny imaginary features off the zone center are purely artifactual. Lack of genuine imaginary modes suggests that monoclinic $AgF_2$ and orthorhombic $CuF_2$ are local minima at the potential energy surface and they could in principle be observed at T→0 K and p→ 0 GPa.

**Table. 3** The list of Γ-point frequencies of the $AgF_2$ and $CuF_2$ compounds in *Pbca* and *P*2$_1$/*c* structures (DFT+U+vdW). Activity of modes in parentheses. Shadowed columns correspond to the translational modes. Frequency in $cm^{-1}$.

| *Pbca* | | | | *P*2$_1$/*c* | | | |
|---|---|---|---|---|---|---|---|
| # | $AgF_2$ [$cm^{-1}$] | $CuF_2$ [$cm^{-1}$] | Symm. (activity) | # | $AgF_2$ [$cm^{-1}$] | $CuF_2$ [$cm^{-1}$] | Symm. (activity) |
| *1* | 478 | 526 | $B_{1g}$ (RAMAN) | *1* | 493 | 560 | $B_g$ (RAMAN) |
| *2* | 475 | 524 | $B_{2g}$ (RAMAN) | *2* | 453 | 490 | $B_u$ (IR) |
| *3* | 473 | 528 | $A_u$ (silent) | *3* | 452 | 500 | $A_u$ (IR) |
| *4* | 467 | 511 | $B_{1u}$ (IR) | *4* | 355 | 389 | $A_g$ (RAMAN) |
| *5* | 464 | 518 | $B_{3u}$ (IR) | *5* | 333 | 373 | $A_u$ (IR) |
| *6* | 450 | 476 | $B_{2u}$ (IR) | *6* | 326 | 357 | $B_u$ (IR) |
| *7* | 369 | 432 | $B_{2u}$ (IR) | *7* | 241 | 294 | $B_g$ (RAMAN) |
| *8* | 347 | 395 | $A_u$ (silent) | *8* | 240 | 263 | $A_g$ (RAMAN) |
| *9* | 332 | 388 | $B_{3g}$ (RAMAN) | *9* | 183 | 250 | $B_u$ (IR) |
| *10* | 329 | 372 | $B_{1u}$ (IR) | *10* | 180 | 254 | $A_u$ (IR) |
| *11* | 325 | 371 | $B_{3u}$ (IR) | *11* | 175 | 235 | $B_g$ (RAMAN) |
| *12* | 324 | 369 | $B_{3g}$ (RAMAN) | *12* | 158 | 220 | $A_u$ (IR) |
| *13* | 307 | 347 | $A_g$ (RAMAN) | *13* | 149 | 192 | $B_u$ (IR) |
| *14* | 274 | 318 | $B_{1g}$ (RAMAN) | *14* | 108 | 145 | $A_u$ (IR) |
| *15* | 273 | 357 | $B_{2g}$ (RAMAN) | *15* | 54 | 75 | $A_g$ (RAMAN) |
| *16* | 240 | 242 | $A_g$ (RAMAN) | *16* | 0 | 0 | $B_u$ |
| *17* | 203 | 267 | $B_{3u}$ (IR) | *17* | 0 | 2 | $A_u$ |
| *18* | 197 | 261 | $B_{1u}$ (IR) | *18* | 1 | 1 | $B_u$ |
| *19* | 191 | 254 | $A_u$ (silent) | | | | |
| *20* | 186 | 283 | $B_{1g}$ (RAMAN) | | | | |
| *21* | 172 | 231 | $A_u$ (silent) | | | | |
| *22* | 172 | 236 | $B_{2u}$ (IR) | | | | |
| *23* | 165 | 203 | $B_{2u}$ (IR) | | | | |
| *24* | 148 | 172 | $B_{3u}$ (IR) | | | | |
| *25* | 148 | 160 | $B_{1u}$ (IR) | | | | |
| *26* | 137 | 157 | $B_{2g}$ (RAMAN) | | | | |
| *27* | 136 | 231 | $B_{3g}$ (RAMAN) | | | | |
| *28* | 114 | 160 | $A_u$ (silent) | | | | |
| *29* | 105 | 126 | $B_{2u}$ (IR) | | | | |
| *30* | 99 | 113 | $B_{3u}$ (IR) | | | | |
| *31* | 90 | 104 | $A_u$ (silent) | | | | |
| *32* | 79 | 180 | $A_g$ (RAMAN) | | | | |
| *33* | 45 | 71 | $B_{1u}$ (IR) | | | | |
| *34* | 0 | 1 | $B_{2u}$ | | | | |
| *35* | 1 | 2 | $B_{1u}$ | | | | |
| *36* | 1 | 1 | $B_{3u}$ | | | | |

On the basis of Group Theory, for the $P2_1/c$ structure (Z=2), among the 18 Γ-point optical modes 9 are IR-active ($5A_u + 4B_u$) and 6 are Raman-active ($3A_g + 3B_g$). For *Pbca* structure Group Theory predicts that 15 modes are IR-active ($5B_{1u} + 5B_{2u} + 5B_{3u}$), 12 are Raman-active ($3A_g + 3B_{1g} + 3B_{2g} + 3B_{3g}$) and 6 are silent ($A_u$). The theoretical results obtained from all theoretical methods used in this work are presented in *Supplementary Information* (Table SI8-SI11, ESI). Here we show theoretical frequencies of the modes, their symmetry and activity, following DFT+U+vdw results only (**Table 3**).

The good agreement between theoretical and experimental values of the mode frequencies reflects the precision of the DFT method (Table SI7, ESI). Moreover, translations (acoustic modes) are computed with +/- 2 $cm^{-1}$ error bar, suggesting that optical frequencies may have a similarly small error.

Raman and IR spectroscopies are very powerful tools in the detection of structural transitions of compounds as they give insight into the local environment of atoms and can be easily equipped with e.g. diamond anvil cell (DAC) used in high-pressure studies. Therefore, analysis of RAMAN and IR spectra may be suitable for disclosure of possible structural transition of $AgF_2$. To anticipate changes on the spectra during such a transition, we will now analyze IR- and RAMAN-active modes for $AgF_2$ and, also, for $CuF_2$ in both structural forms. We note that the evolution of RAMAN spectra of $CuF_2$ upon pressure, has recently been employed to detect $P2_1/c$ -> *Pbca* transition[11]. Thus, this work completes previous research by including data of IR-active modes for this compound, in its both structural forms.

Based on **Table 3**, there are clear differences in the fundamental phonons, between $AgF_2$ and $CuF_2$, in two polymorphic forms. The symmetry-related modes (e.g. $B_{1g}$ in both compounds in orthorhombic form) have higher frequencies in case of copper difluoride. In the simplest harmonic approximation, the vibration energy is proportional to the bonding strength (force constant) and inversely proportional to the mass of interacting atoms (reduced mass). Although, the M-F bond is more covalent in $AgF_2$ than in $CuF_2$[18,39,40], the reduced mass seems to be a dominant factor of the aforementioned changes in the frequencies of modes; mass of Cu constitutes only ca. 60% of atomic mass of Ag.

Regarding the experimentally-related issue, transition from monoclinic to orthorhombic structure of $AgF_2$ should be accompanied with reduction of the number of band appearing in both IR and RAMAN spectra, from 27 to 15 optically active modes. Additionally, *Pbca* -> $P2_1/c$ transition should be reflected in the changes of the frequencies of symmetry-related modes. For instance, the sudden change of 15-18 $cm^{-1}$ is expected, between symmetry-related $B_{1g}$ (478 $cm^{-1}$) and $B_{1g}$ (475 $cm^{-1}$) modes for *Pbca* and $B_g$ (493 $cm^{-1}$) for $P2_1/c$ polymorphs. This trend of mode stiffening is more general, as the ZPE value is predicted to by higher by 7 meV/FU (**Table 4**) for $P2_1/c$ form, than that, for the ambient pressure form of silver difluoride. This emerges partially from the changes in the average value of associated bond lengths. Based on the theoretical results, the shortening by approx. 0.1% and 0.9%, for equatorial and axial M-F bonds, respectively, following *Pbca* -> $P2_1/c$ transformation, should indeed take place for $AgF_2$. Same trend holds for $CuF_2$.

### *1.3. Energetic and thermodynamic stability of both polymorphs*

Comparing ground state energies of binary fluorides in $P2_1/c$ and *Pbca* systems (including zero-point energies) and volume changes (dV) arising from their structure modification, we may estimate the relative stability of both polymorphs as well as transition pressure of $P2_1/c$ -> *Pbca* transformation for both compounds. In doing so we follow equation $p = \frac{dE}{dV}$ (1) and assume so called common tangent approximation. The results obtained with the state of the art DFT+U+vdW method are listed in **Table 4** (for all methods tested *cf.* Table SI3-SI4 in ESI).

Copper difluoride in the *Pbca* structure is uphill by 80 meV per FU (including ZPE) as compared to monoclinic form; this explains persistence of the latter in experiment at ambient (p,T) conditions. However, the volume of *P*nma form is by 1.65-1.76 Å$^3$ smaller than in case of the monoclinic form, suggesting that orthorhombic form could be achieved at elevated pressure. Our computed transition pressure of 7.3 GPa is only slightly underestimated (experimental value is 1.7 GPa larger[11]).

**Table 4.** Ground state energies ($E_{GS}$), volumes of unit cell (V), zero-point energies (ZPE), transition pressure (p) including zero-point energy ($p_{ZPE}$), transition temperature from ambient pressure structure without (T) and with zero-point energy ($T_{ZPE}$). Data within DFT+U+vdW method. Results from all methods presented in *Supplementary Information,* Table. SI4-SI5

| **$AgF_2$** | **$E_{GS}$/FU [eV]** | **V/FU [Å$^3$]** | **ZPE [eV]** | **p [GPa]** | **$p_{ZPE}$ [GPa]** | **T [K]** | **$T_{ZPE}$ [K]** |
|---|---|---|---|---|---|---|---|
| *$P2_1/c$* | -9.224 | 40.995 | 0.121 | -2.8 | -2.4 | 573 | 480 |
| *Pbca* | -9.273 | 38.211 | 0.129 | | | | |
| **$CuF_2$** | **$E_{GS}$/FU [eV]** | **V/FU [Å$^3$]** | **ZPE [eV]** | **p [GPa]** | **$p_{ZPE}$ [GPa]** | **T [K]** | **$T_{ZPE}$ [K]** |
| *$P2_1/c$* | -11.626 | 32.068 | 0.143 | 6.3 | 7.3 | 805 | 933 |
| *Pbca* | -11.557 | 30.311 | 0.154 | | | | |

In case of $AgF_2$, relative stability of two structures depends on the method used (see ESI). For DFT+U+vdW standard the *Pbca* form is more stable than monoclinic form by 49 meV (including ZPE). On the other hand, unit cell volume of $AgF_2$ in *Pbca* form is ca. 1.7 Å$^3$ smaller than in $P2_1/c$. Therefore, according to the equation 1, calculated pressure of transition must be negative. Here, we obtain the value of ca. -2.4 GPa with DFT+U+vdW (**Table 4**). Since the ambient (p,T) form of $AgF_2$ is orthorhombic (*Pbca*), we believe that employment of van der Waals functional is necessary for a proper description of the relative stability. Nevertheless, the formally negative value of transition pressure, whie unphysical, may to some extent be mimicked by negative strain (e.g. in epitaxial deposition techniques). On the other hand, one may estimate temperature needed for structural transition to take place. In case of silver difluoride, the estimated temperature for *Pbca* -> $P2_1/c$ structural transformation is equal ca. 450 K (with ZPE correction). We will return to this finding below.

### *1.4. Electronic and magnetic properties*

The **Fig. 5** presents electronic density of states for copper and silver difluorides in two structures: *Pbca* and $P2_1/c$. The value of fundamental band gap, calculated for $AgF_2$ (*Pbca*) is consistent within three methods used: DFT+U+vdW and HSE06, and equals ca. 2.17-2.18 eV (**Table 4**). However, for SCAN method yields a severly underestimated value of BG of 0.52 eV. Since experimentally determined charge-transfer gap according to the ZSA scheme[41] is ca. 3.4 eV[19], we conclude that even HSE06 method underestimates the gap. For monoclinic form of $AgF_2$ ($P2_1/c$), calculated charge-transfer gap is approx. 1.94-1.96 eV (within DFT+U+vdW and HSE06). These values are 11 % smaller than the gap computed for orthorhombic form so, given error cancelling, we expect that experimentally determined BG for monocilinic form would indeed be smaller than that for orthorhombic form (89% of 3.4 eV yields ca. 3.0 eV).

The changes of the band gap value with structural transformation are also observed in case of $CuF_2$. For the ambient pressure form of copper difluoride, the DFT+U+vdw method gives BG of 4.46 eV, while HSE06 predicts a somewhat smaller value of 4.14 eV. To the best of our knowledge, there is no experimental estimate of the bang gap value for $CuF_2$. However, our theoretical results are in line with previously reported calculations by Miller and Botana[40]. It may be expected that the band gap for more ionic $CuF_2$ should indeed be larger than for more covalent $AgF_2$. Again, as for $AgF_2$, calculations predict larger gap for orthorhombic than for the monoclinic form (by 4% and 7%, following DFT+U+vdw and HSE06 results, respectively). To verify this result, high-pressure form of $CuF_2$ should first be quenched to ambient (p,T) conditions and its band gap examined.

The result obtained for $AgF_2$ agree qualitatively with the Maximum Hardness Principle (MHP) by Pearson[42] as a broader gap (*Pbca*) form is more stable (at p→ 0 atm, T→ 0 K) than a narrower-gap *P*$2_1$/c form. Interestingly, contradictory observation can be made for $CuF_2$.

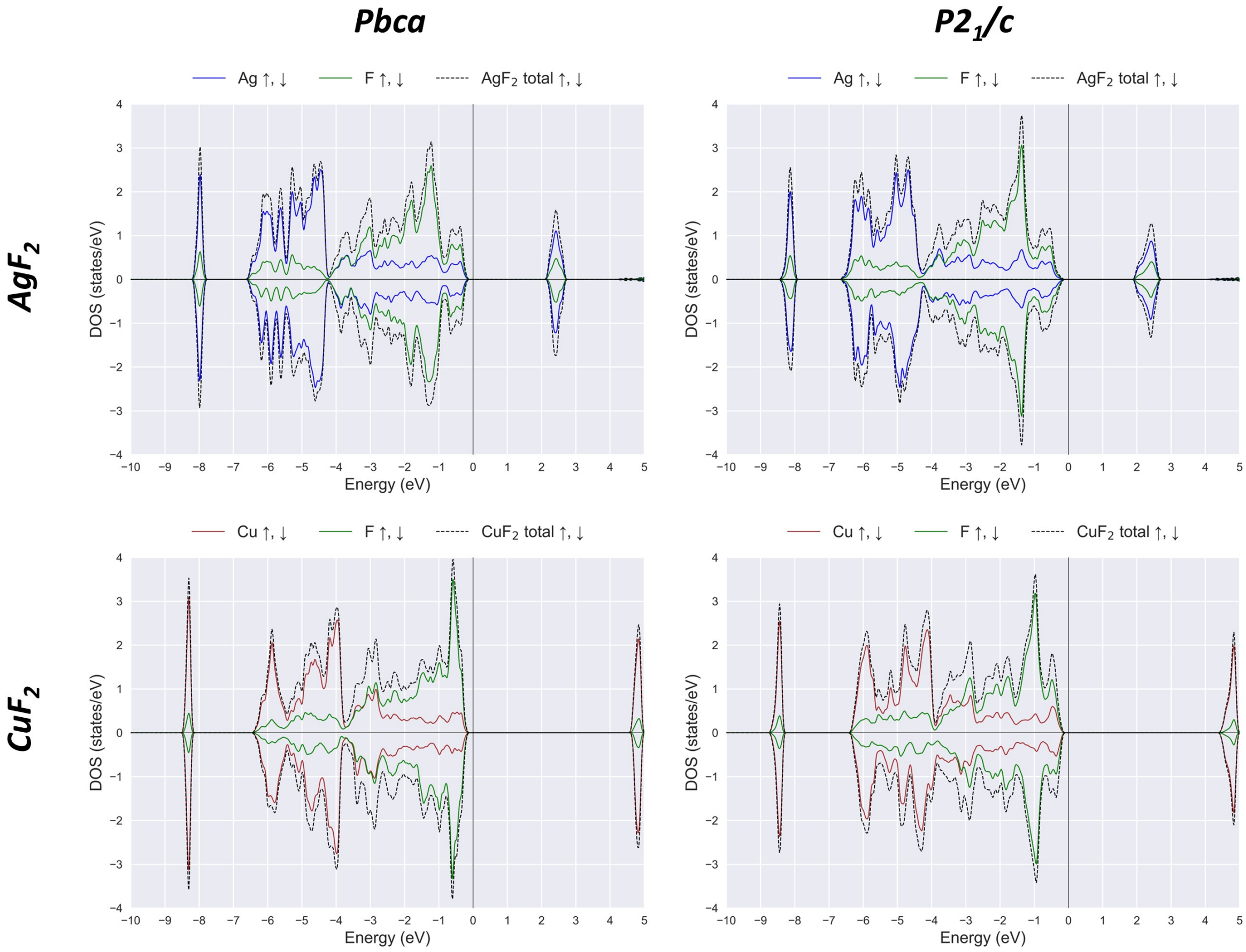

**Fig. 5** Atom-resolved eDOS for $AgF_2$ (upper) and $CuF_2$ (lower) in *Pbca* and *P*$2_1$/c systems calculated within DFT+U+vdW method. For eDOS from HSE06 method, please check **Fig. SI1** in **ESI**.

The inspection of density of states (DOS presented on **Fig. 5)** indicates, that between *P*$2_1$/*c* and *Pbca* forms of $AgF_2$, there are some differences regarding contributions from silver and fluorine states to conduction and valence bands. In case of orthorhombic form of silver difluoride (ambient pressure), the major contribution to valence band (at the vicinity of Fermi level) comes from $F_p$ states, whereas, the conduction band (i.e. upper Hubbard band, UHB) is mainly formed by $Ag_d$ states. In case of monoclinic form, one may note a stronger admixture of Ag and F states in both bands. This indicates increased Ag-F bond covalence, in agreement to the changes of bond lengths and optical phonon frequencies. This tendency is even more evident in case of $CuF_2$ (**Fig. 5**, bottom part). Here, again, monoclinic form shows greater hybridization of $Cu_d$-$F_p$ states than the orthorhombic one.

The width of Upper Hubbard Band (conduction band) is another important parameter related to possible electron doping to both compounds.[22,43] Our results indicate, that in both compounds, the UHB is slightly widened in *P*$2_1$/*c* structure as compared to *P*bca one. For this reason, $AgF_2$ in monoclinic structure might be more feasible to dope, than in its ambient (p,T) form[21].

As $AgF_2$ and $CuF_2$ are two-dimensional antiferromagnets( for $AgF_2$ the $T_N$ = 163K[44], whereas for $CuF_2$, $T_N$ = 69K[45]), the strength of magnetic superexchange, $J_{2D}$, is of interest. Experimentally determined $J_{2D}$ value for $AgF_2$ is equal ca. -70 meV[17] (negative sign indicates antiferromagnetic interaction). The SCAN predicts very similar value: -67.3 meV; in DFT+U method, the superexchange value is twice as low (**Table 5**). In case of $CuF_2$, limited information is available on the experimentally determined value of the intra-sheet magnetic superexchange constant. To the best of our knowledge, there is only one experimental paper, where authors claim the magnitude of $J_{2D}$ to be -2.9 meV[45], on the basis of magnetic measurements. However, sophisticated computational method predicts it to be *ca.* -11 meV[46]. Our calculations are not so far from the latter; DFT+U+vdW predicts value of -8.2 meV. On the other hand, SCAN and HSE06 methods predict higher absolute value of $J_{2D}$ i.e. -21.6 meV and -18.3 meV, respectively (**Table 5**). The exact SE value in DFT+U methods, strongly depends on the magnitude of the U and J parameters used[25,40]. Therefore, regarding results from DFT+U+vdW methods, we mainly focus on the changes of $J_{2D}$ values between two polymorphs.

**Table 5.** Band gaps, magnetic moments and $J_{2D}$ values for $AgF_2$ and $CuF_2$ in *Pbca* and $P2_1/c$ structures, obtained using various calculation methods. Negative values of $J_{2D}$ indicate antiferromagnetic interaction.

| Method | System | $AgF_2$ | | | $CuF_2$ | | |
|---|---|---|---|---|---|---|---|
| | | **BG [eV]** | **$\pm\mu_{AFM}$** | **$J_{2D}$ [meV]** | **BG [eV]** | **$\pm\mu_{AFM}$** | **$J_{2D}$ [meV]** |
| DFT+U | *Pbca* | 2.18 | 0.66 | -32.7 | 4.63 | 0.88 | -6.6 |
| | $P2_1/c$ | 1.95 | 0.67 | -40.2 | 4.46 | 0.89 | -8.2 |
| DFT+U+vdW | *Pbca* | 2.17 | 0.67 | -31.9 | 4.64 | 0.88 | -6.6 |
| | $P2_1/c$ | 1.94 | 0.67 | -39.4 | 4.46 | 0.89 | -8.2 |
| SCAN | *Pbca* | 0.52 | 0.52 | -67.3 | 1.58 | 0.75 | -18.8 |
| | $P2_1/c$ | 0.35 | 0.52 | -81.0 | 1.21 | 0.76 | -21.6 |
| HSE06 | *Pbca* | 2.18 | 0.60 | -54.7 | 4.46 | 0.81 | -15.6 |
| | $P2_1/c$ | 1.96 | 0.60 | -63.2 | 4.14 | 0.81 | -18.3 |

Calculations reveal enhancement of $J_{2D}$ absolute values in $AgF_2$ in $P2_1/c$ forms by 13%-19%, depending on the method (**Table 5**, **Table SI5**). This trend is also valid in the case of $CuF_2$ (13%-20%). Such changes of the $J_{2D}$ value are strictly related to the changes of M-M distances, and linked to that, opening of the F-M-F angle driven by structural transformation *P*bca -> $P2_1$/c. According to the GKA[47–49] rules more effective overlapping of the orbitals should be expected with flattening of the F-M-F angle, and so, stronger hybridization of M-F orbitals leads to higher absolute value of antiferromagnetic $J_{2D}$. Since the $J_{2D}$ value is related to the maximum experimental superconducting critical temperature in doped cuprates,[50] synthesis of monoclinic $AgF_2$ precursor is worthwhile.

Another interesting aspect is the relationship between charge transfer energy and the $J_{2D}$ value. For charge transfer insulators, hopping of electron from *d* orbital of transition metal site to another and back, engages *p* of nonmetal, thus J is proportional to 1/ $(\Delta_{CT})^2$[47–49,51].Indeed, our results show that larger $|J_{2D}|$ stems from smaller $\Delta_{CT}$. This dependence is consistent within all methods employed (**Table 5**, *cf.* **ESI**), for both compounds in *P*bca and $P2_1/c$ structures.

## Conclusions

In this paper, employing various theoretical methods, we investigated the structural, electronic and magnetic properties of the known, high-pressure orthorhombic form of $CuF_2$ (*P*bca) and not-yet achieved monoclinic form of $AgF_2$ ($P2_1/c$), and compared them with those for ambient (p,T) polymorphs. Both fluorides are dynamically stable in the monoclinic structure and they exhibit higher $|J_{2D}|$ value, smaller charge-transfer gap and wider UHB, comparing to orthorhombic one. These findings are especially important in case of $AgF_2$, which is perceived as analogue of cuprates. Stabilization of the monoclinic form of silver(II) difluoride ($P2_1/c$) may be crucial in order to achieving long-awaited doping of this compound.

Based on our theoretical results, we claim that the desired structural transition of $AgF_2$ to monoclinic form appears to be achievable in two possible ways (aside from partial Cu-doping[25]): 1) at negative strain conditions[43], or 2) upon heating to T> 480 K and quenching. Regarding 1), interatomic

Ag-Ag distance in $AgF_2$ layer may be controlled by lattice parameters of substrate, and thus ensuring proper strain for crystal growth of the $P2_1/c$ polytype. Regarding 2), heating to 480 K should be accompanied by fluorine gas overpressure needed to prevent thermal decomposition. Regretfully, so far, such sudden thermal quenching experiments were not yet performed for $AgF_2$.

**Acknowledgements**

W.G. is grateful to the Polish National Science Center (NCN) for project Maestro (2017/26/A/ST5/00570). Research was carried out with the use of CePT infrastructure financed by the European Union – the European Regional Development Fund within the Operational Programme "Innovative economy" for 2007-2013 (POIG.02.02.00-14-024/08-00). Quantum mechanical calculations were performed using supercomputing resources of the ICM UW (projects SAPPHIRE GA83-34). Authors appreciate valuable comments from prof. Jose Lorenzana (CNR and La Sapienza, Rome, Italy).

**SUPPLEMENTARY INFORMATION**

# Polymorphism of two-dimensional antiferromagnets, $AgF_2$ and $CuF_2$

**Daniel Jezierski* and Wojciech Grochala***

*d.jezierski@cent.uw.edu.pl, w.grochala@cent.uw.edu.pl

*Center of New Technologies, University of Warsaw, 020289 Warsaw Poland*

## List of contents

## SI1. Pressure-induced phase transitions in $TMF_2$

Main types of structures:

1) **Rutile:**
   a. $TiO_2$ (*$P4_2/mnm$*) | tetragonal
   b. $CaCl_2$ (*Pnnm*) | orthorhombic
   c. distorted-rutile (*$P2_1/c$*) |monoclinic
2) **Fluorite**
   a. HP-$PdF_2$ ($Pa\bar{3}$) | cubic
   b. $CaF_2$ ($Fm\bar{3}m$) | cubic
   c. distorted-fluorite (*Pbca*) |orthorhombic
   d. noncentrosymmetric ($Pca2_1$) | orthorhombic
   e. nanotubes (*Pbcn*) | orthorhombic
3) **Cotunnite**
   a. α-$PbCl_2$ (*Pnma*) | orthorhombic
4) **Hexagonal**
   a. $Ni_2In$ ($P6_3/mmc$) | hexagonal

**$MnF_2$**: $P4_2/mnm \xrightarrow{7.2\ GPa} Pbca$ ($SrI_2$) $\xrightarrow{33\ GPa} Pnma$ [1]

**$FeF_2$**: $P4_2/mnm \xrightarrow{4.5\ GPa} Pnnm \xrightarrow{25\ GPa}$ hexagonal phase (probably $P6_3/mmc$) [2,3] | no SG or Wyckoff position

**$CoF_2$**: $P4_2/mnm \xrightarrow{3.6\ GPa} Pnnm \xrightarrow{8\ GPa}$ phase coexistence ($Pnnm + Pa\bar{3} + Pbca$) $\xrightarrow{15\ GPa} Fm\bar{3}m \xrightarrow{44\ GPa} Pnma$ [4]

**$NiF_2$**: $P4_2/mnn \xrightarrow{1.8\ GPa} Pnnm$[5] $\xrightarrow{30\ GPa} C2/m$[6]

**$PdF_2$**: $P4_2/mnn \xrightarrow{1.5\ GPa} Pa\bar{3}$ [7]

***$CuF_2$**: $P2_1/c \xrightarrow{9\ GPa} Pbca \xrightarrow{30\ GPa} Pca2_1 \xrightarrow{72\ GPa}$ Pnma [8]

**$AgF_2$**: $Pbca \xrightarrow{4.9\ GPa} Pca2_1 \xrightarrow{14\ GPa} Pbcn \xrightarrow{36.2\ GPa}$ Pnma [9]

**$ZnF_2$**: $P4_2/mnn \xrightarrow{4.5\ GPa} Pnnm$[10] $\xrightarrow{10\ GPa} Pa\bar{3} \xrightarrow{30\ GPa} Pca2_1 \xrightarrow{44\ GPa} Pnma$[11]

$P4_2/mnn \xrightarrow[673K]{5.4\ GPa} Pnnm \xrightarrow[623K]{15.3\ GPa} Pa\bar{3}$[12]

**$CdF_2$**: $Fm\bar{3}m \xrightarrow{7\ GPa} Pnma$[13] $\xrightarrow{87\ GPa} P6_3/mmc$ [14]

**$HgF_2$**: $Fm\bar{3}m \xrightarrow{4.7\ GPa} Pnma$ [15]

blue colour : DFT+U predictions

*$CuF_2$ compressed only up to 29.4 GPa. Theoretical prediction indicates, that the next structural transition should be around 30 GPa: *Pbca* -> $Pca2_1$ (non-centrosymmetric).

## SI2. Bond lengths for in optimized $P2_1/c$ and *Pbca* forms of $AgF_2$ and $CuF_2$

**Table SI1.** Ag-F and Ag-Ag distances in $AgF_2$ compound in $P2_1/c$ and *Pbca.* Values in Å, based on fully optimized structures within all methods employed.

| | d Ag-F [Å] | | | | d Ag-Ag [Å] | | |
|---|---|---|---|---|---|---|---|
| ***Pbca*** | **in layer** | | **apical** | **secondary** | **in layer** | **out layer** | **secondary** |
| *DFT+U* | 2.055 | 2.052 | 2.529 | 3.137 | 3.694 | 3.819 | 3.962 |
| *DFT+U+vdW* | 2.050 | 2.047 | 2.477 | 3.180 | 3.680 | 3.774 | 3.914 |
| *SCAN* | 2.087 | 2.089 | 2.570 | 3.135 | 3.798 | 3.850 | 3.998 |
| *HSE06* | 2.067 | 2.069 | 2.612 | 3.180 | 3.759 | 3.872 | 4.025 |
| **$P2_1/c$** | **in layer** | | **apical** | **secondary** | **in layer** | **out layer** | **secondary** |
| *DFT+U* | 2.043 | 2.060 | 2.489 | 3.627 | 3.721 | 3.638 | 4.007 |
| *DFT+U+vdW* | 2.028 | 2.062 | 2.457 | 3.569 | 3.700 | 3.655 | 3.944 |
| *SCAN* | 2.075 | 2.095 | 2.546 | 3.687 | 3.810 | 3.743 | 4.025 |
| *HSE06* | 2.061 | 2.074 | 2.572 | 3.683 | 3.775 | 3.722 | 4.063 |

**Table SI2.** Cu-F and Cu-Cu distances in $CuF_2$ compound in *Pbca* and $P2_1/c$ form. Values in Å, based on fully optimized structures within all methods employed.

| | d Cu-F [Å] | | | | dCu-Cu [Å] | | |
|---|---|---|---|---|---|---|---|
| **Pbca** | **in layer** | | **apical** | **secondary** | **in layer** | **out layer** | **secondary** |
| *DFT+U* | 1.897 | 1.902 | 2.321 | 2.963 | 3.410 | 3.535 | 3.637 |
| *DFT+U+vdW* | 1.898 | 1.894 | 2.294 | 2.956 | 3.398 | 3.512 | 3.612 |
| *SCAN* | 1.904 | 1.910 | 2.403 | 3.020 | 3.442 | 3.590 | 3.712 |
| *HSE06* | 1.897 | 1.901 | 2.421 | 3.012 | 3.435 | 3.596 | 3.718 |
| **$P2_1/c$** | **in layer** | **apical** | **secondary** | **in layer** | **out layer** | **secondary** | **in layer** |
| *DFT+U* | 1.895 | 1.910 | 2.210 | 3.434 | 3.475 | 3.192 | 3.672 |
| *DFT+U+vdW* | 1.886 | 1.910 | 2.192 | 3.413 | 3.464 | 3.185 | 3.646 |
| *SCAN* | 1.899 | 1.919 | 2.302 | 3.442 | 3.487 | 3.302 | 3.740 |
| *HSE06* | 1.898 | 1.910 | 2.307 | 3.443 | 3.482 | 3.296 | 3.741 |

## SI3. Thermodynamic data and superexchange constants for $AgF_2$ and $CuF_2$ in both structural forms

**Table SI3.** Ground state energy (GS energy), zero-point energies (ZPE), volumes of the unit cell (V), band gap values (BG), energy difference between $\mathbf{AgF_2}$ in two systems (dE), volume difference of $AgF_2$ in two systems (dV) and the equilibrium pressure (p) for two $AgF_2$ systems – *Pbca* and *P21/c*. dE = $E_{P21/c}$ - $E_{Pbca}$ [meV], dV= $V_{Pbca}$ - $V_{P21/c}$.

| Structure | E/FU [ev] | V/FU [Å³] | ZPE/FU [eV] | E with ZPE [ev] | dE [meV] | dE $_{ZPE}$ [meV] | dV | p [GPa] | $p_{ZPE}$ [GPa] | BG [eV] |
|---|---|---|---|---|---|---|---|---|---|---|
| *DFT+U, U = 8 eV* | | | | | | | | | | |
| *Pbca*<br>*P21/c* | -8.806<br>-8.808 | 39.233<br>42.006 | 0.125<br>0.119 | -8.681<br>-8.689 | -2.338 | -8.234 | -2.774 | 0.1 | 0.5 | 2.18<br>1.95 |
| *DFT + U + vdW, U = 8 eV* | | | | | | | | | | |
| *Pbca*<br>*P21/c* | -9.273<br>-9.224 | 38.211<br>40.995 | 0.129<br>0.121 | -9.144<br>-9.103 | 49.383 | 41.404 | -2.785 | -2.8 | -2.4 | 2.17<br>1.94 |
| *SCAN* | | | | | | | | | | |
| *Pbca*<br>*P21/c* | -127.382<br>-127.345 | 41.145<br>44.176 | 0.123<br>0.102 | -127.259<br>-127.243 | 36.635 | 15.880 | -3.031 | -1.9 | -0.8 | 0.52<br>0.35 |
| *HSE06* | | | | | | | | | | |
| *Pbca*<br>*P21/c* | -13.491<br>-13.488 | 41.071<br>43.850 | 0.126<br>0.120 | -13.366<br>-13.368 | 3.707 | -2.167 | -2.779 | -0.2 | 0.1 | 2.18<br>1.96 |

**Table SI4.** Ground state energy (GS energy), zero-point energies (ZPE), volumes of unit cell (V), band gap values (BG), energy difference between **$CuF_2$** in two systems (dE), volume difference of $AgF_2$ in two systems (dV) and the equilibrium pressure (p) for two $CuF_2$ systems – *Pbca* and *P21/c*. dE = $E_{Pbca}$ - $E_{P21/c}$ [meV], dV= $V_{P21/c}$ - $V_{Pbca}$

| Structure | E/FU [ev] | V/FU [Å³] | ZPE/FU [eV] | E with ZPE [ev] | dE [meV] | dE $_{ZPE}$ [meV] | dV | p [GPa] | $p_{ZPE}$ [GPa] | BG [eV] |
|---|---|---|---|---|---|---|---|---|---|---|
| | | | | *DFT+U, U = 10 eV* | | | | | | |
| *Pbca* | -11.165 | 30.803 | 0.150 | -11.015 | 102.837 | 112.432 | 1.706 | 9.7 | 10.6* | 4.63 |
| *P21/c* | -11.268 | 32.510 | 0.140 | -11.128 | | | | | | 4.46 |
| | | | | *DFT + U + vdW, U = 10 eV* | | | | | | |
| *Pbca* | -11.557 | 30.311 | 0.154 | -11.403 | 69.358 | 80.423 | 1.756 | 6.3 | 7.3 | 4.64 |
| *P21/c* | -11.626 | 32.068 | 0.143 | -11.483 | | | | | | 4.46 |
| | | | | *SCAN* | | | | | | |
| *Pbca* | -80.825 | 32.154 | 0.145 | -80.680 | 40.923 | 52.928 | 1.763 | 3.7 | 4.8 | 1.58 |
| *P21/c* | -80.866 | 33.917 | 0.133 | -80.733 | | | | | | 1.21 |
| | | | | *HSE06* | | | | | | |
| *Pbca* | -16.241 | 32.073 | 0.154 | -16.087 | 65.676 | 76.914 | 1.650 | 6.4 | 7.5 | 4.46 |
| *P21/c* | -16.307 | 33.724 | 0.143 | -16.164 | | | | | | 4.14 |

* transition at 9GPa reported experimentally by Kurzydłowski[8]

**Table SI5.** $J_{2D}$[meV] value for $AgF_2$ and $CuF_2$ in two types of structure: *P21/c* and *Pbca* calculated using four methods: DFT+U, DFT+U+vDW, SCAN, HSE06. F-Ag-F angles for in-layer plaquettes.

| *Method* | $J_{2D}$ [mev] | | | | | | | | | |
|---|---|---|---|---|---|---|---|---|---|---|
| | *$AgF_2$* | | | | | *$CuF_2$* | | | | |
| | *Pbca* | *F-Ag-F angle [°]* | *P21/c* | *F-Ag-F angle [°]* | *$J_{Pbca}/J_{P21/c}$* | *Pbca* | *F-Ag-F angle [°]* | *P21/c* | *F-Ag-F angle [°]* | *$J_{Pbca}/J_{21/c}$* |
| DFT+U | -32.7 | 128.2 | -40.7 | 130.1 | 0.81 | -6.6 | 127.7 | -8.2 | 132.0 | 0.81 |
| DFT+U+vdW | -31.9 | 127.9 | -39.4 | 129.5 | 0.81 | -6.6 | 127.3 | -8.2 | 131.7 | 0.80 |
| SCAN | -67.3 | 130.8 | -81.0 | 132.0 | 0.83 | -18.8 | 128.9 | -21.6 | 131.9 | 0.87 |
| HSE06 | -52.0 | 130.6 | -59.1 | 131.8 | 0.87 | -15.6 | 129.4 | -18.3 | 132.2 | 0.85 |

# SI4. Optimized structure following imaginary mode

**Table SI6.** Parameters of unit cells obtained after optimization of the specific displacement, following imaginary mode. The symmetry of the optimized structure (marked as i-mode) was found using Materials Studio software.

| | $CuF_2$ | | $AgF_2$ | |
|---|---|---|---|---|
| **Parameters** | i-mode ($P2_1/c$) | *P21/c* | i-mode ($P2_1/c$) | *P21/c* |
| **a [Å]** | 4.543 | 3.192 | 4.871 | 3.638 |
| **b [Å]** | 4.504 | 4.505 | 4.775 | 4.777 |
| **c [Å]** | 4.800 | 5.293 | 6.428 | 5.706 |
| **alpha [°]** | 90.000 | 90.000 | 90.000 | 90.000 |
| **beta [°]** | 146.784 | 122.091 | 145.848 | 122.091 |
| **gamma [°]** | 90.000 | 90.000 | 90.000 | 90.000 |
| **V/FU [Å³]** | 32.503 | 32.510 | 41.965 | 42.006 |
| **E/FU [eV]** | -11.268 | -11.268 | -8.808 | -8.808 |
| $E_{diff}$ ($E_{i-mode}$ - $E_{P21/c}$) [meV] | **-0.113** | | **-0.105** | |

Imaginary mode for $AgF_2$ and $CuF_2$, both in the *P21/c* structure, with minimum around Gamma point q = (-0.04, 0.04, -0.04). Frequency of minimum point: -8.387 $cm^{-1}$ and -10.464 $cm^{-1}$ for $AgF_2$ and $CuF_2$, respectively.

Optimized structures are ca. -0.1 meV lower in energy from parent structures, which is v=within error bar of the DFT method, and strongly suggests that imaginary character of the phonon is **artifactual**.

## SI5. Vibrational frequencies for $AgF_2$/$CuF_2$ in $P2_1/c$ and *Pbca* forms

**Table SI7.** Phonon frequencies for $AgF_2$ and $CuF_2$ in *Pbca* systems (DFT+U). The last three bands correspond to the translations. For $CuF_2$ theoretical frequencies were assigned to the experimental values, with classification of the modes based on the relative intensity: shoulder (sh), very strong (vs), strong (s), medium (m) and weak (w).

| Pbca | | | | $P2_1/c$ | | | | |
|---|---|---|---|---|---|---|---|---|
| # | $AgF_2$* | $CuF_2$ | Symm. | # | $AgF_2$ | $CuF_2$ | Symm. | EXP $CuF_2$ only* |
| *1* | 472 | 523 | $B_{1g}$ (RAMAN) | *1* | 486 | 551 | $B_g$ (RAMAN) | 556 (w) |
| *2* | 469 | 520 | $B_{2g}$ (RAMAN) | *2* | 450 | 493 | $A_u$ (IR) | 508 (sh) |
| *3* | 465 | 523 | $A_u$ (silent) | *3* | 447 | 482 | $B_u$ (IR) | 483 (vs) |
| *4* | 457 | 504 | $B_{1u}$ (IR) | *4* | 341 | 380 | $A_g$ (RAMAN) | 355 (m) |
| *5* | 456 | 513 | $B_{3u}$ (IR) | *5* | 322 | 363 | $A_u$ (IR) | 360 (sh) |
| *6* | 441 | 473 | $B_{2u}$ (IR) | *6* | 319 | 351 | $B_u$ (IR) | 354 (s) |
| *7* | 361 | 427 | $B_{2u}$ (IR) | *7* | 242 | 258 | $A_g$ (RAMAN) | 254 (s) |
| *8* | 334 | 385 | $A_u$ (silent) | *8* | 237 | 292 | $B_g$ (RAMAN) | 293 (vs) |
| *9* | 330 | 386 | $B_{3g}$ (RAMAN) | *9* | 183 | 246 | $B_u$ (IR) | 243 (s) |
| *10* | 320 | 366 | $B_{1u}$ (IR) | *10* | 180 | 232 | $B_g$ (RAMAN) | 221 (s) |
| *11* | 318 | 366 | $B_{3u}$ (IR) | *11* | 176 | 247 | $A_u$ (IR) | 246 (s) |
| *12* | 313 | 360 | $B_{3g}$ (RAMAN) | *12* | 167 | 220 | $A_u$ (IR) | 220 (sh) |
| *13* | 308 | 349 | $A_g$ (RAMAN) | *13* | 145 | 188 | $B_u$ (IR) | 190 (s) |
| *14* | 264 | 312 | $B_{1g}$ (RAMAN) | *14* | 103 | 140 | $A_u$ (IR) | 136 (m) |
| *15* | 263 | 345 | $B_{2g}$ (RAMAN) | *15* | 48 | 84 | $A_g$ (RAMAN) | n.d. |
| *16* | 234 | 238 | $A_g$ (RAMAN) | *16* | 0 | 1 | $B_u$ | |
| *17* | 195 | 261 | $B_{3u}$ (IR) | *17* | 1 | 2 | $A_u$ | |
| *18* | 191 | 255 | $B_{1u}$ (IR) | *18* | 1 | 2 | $B_u$ | |
| *19* | 188 | 248 | $A_u$ (silent) | | | | | |
| *20* | 181 | 274 | $B_{1g}$ (RAMAN) | | | | | |
| *21* | 168 | 229 | $A_u$ (silent) | | | | | |
| *22* | 166 | 236 | $B_{2u}$ (IR) | | | | | |
| *23* | 158 | 198 | $B_{2u}$ (IR) | | | | | |
| *24* | 146 | 167 | $B_{3u}$ (IR) | | | | | |
| *25* | 139 | 154 | $B_{1u}$ (IR) | | | | | |
| *26* | 136 | 153 | $B_{2g}$ (RAMAN) | | | | | |
| *27* | 127 | 217 | $B_{3g}$ (RAMAN) | | | | | |
| *28* | 102 | 157 | $A_u$ (silent) | | | | | |
| *29* | 97 | 126 | $B_{2u}$ (IR) | | | | | |
| *30* | 91 | 113 | $B_{3u}$ (IR) | | | | | |
| *31* | 75 | 91 | $A_u$ (silent) | | | | | |
| *32* | 69 | 66 | $A_g$ (RAMAN) | | | | | |
| *33* | 35 | 54 | $B_{1u}$ (IR) | | | | | |
| *34* | 0 | 1 | $B_{2u}$ | | | | | |
| *35* | 0 | 1 | $B_{1u}$ | | | | | |
| *36* | 1 | 2 | $B_{3u}$ | | | | | |

* experimental data published by Gawraczyński et.all, PNAS 2019[16], ** for RAMAN data taken from Kurzydłowski[8], for IR own results.

**Table SI8.** Phonon frequencies for $AgF_2$ in *Pbca* system for all methods. The last three bands correspond to the translations.

| # | DFT+U cm$^{-1}$ | DFT+U Symm. | DFT+U+vdw cm$^{-1}$ | DFT+U+vdw Symm. | SCAN cm$^{-1}$ | SCAN Symm. | HSE06 cm$^{-1}$ | HSE06 Symm. |
|---|---|---|---|---|---|---|---|---|
| *1* | 472 | $B_{1g}$ | 478 | $B_{1g}$ | 476 | $A_u$ | 480 | $B_{1g}$ |
| *2* | 469 | $B_{2g}$ | 475 | $B_{2g}$ | 468 | $B_{3u}$ | 477 | $B_{2g}$ |
| *3* | 465 | $A_u$ | 473 | $A_u$ | 467 | $B_{1u}$ | 475 | $A_u$ |
| *4* | 457 | $B_{1u}$ | 467 | $B_{1u}$ | 451 | $B_{2u}$ | 469 | $B_{1u}$ |
| *5* | 456 | $B_{3u}$ | 464 | $B_{3u}$ | 414 | $B_{1g}$ | 467 | $B_{3u}$ |
| *6* | 441 | $B_{2u}$ | 450 | $B_{2u}$ | 411 | $B_{2g}$ | 458 | $B_{2u}$ |
| *7* | 361 | $B_{2u}$ | 369 | $B_{2u}$ | 343 | $B_{2u}$ | 355 | $B_{2u}$ |
| *8* | 334 | $A_u$ | 347 | $A_u$ | 326 | $B_{3g}$ | 343 | $B_{3g}$ |
| *9* | 330 | $B_{3g}$ | 332 | $B_{3g}$ | 317 | $A_u$ | 329 | $A_g$ |
| *10* | 320 | $B_{1u}$ | 329 | $B_{1u}$ | 311 | $A_g$ | 325 | $A_u$ |
| *11* | 318 | $B_{3u}$ | 325 | $B_{3u}$ | 304 | $B_{1u}$ | 318 | $B_{3u}$ |
| *12* | 313 | $B_{3g}$ | 324 | $B_{3g}$ | 302 | $B_{3g}$ | 318 | $B_{1u}$ |
| *13* | 308 | $A_g$ | 307 | $A_g$ | 301 | $B_{3u}$ | 306 | $B_{3g}$ |
| *14* | 264 | $B_{1g}$ | 274 | $B_{1g}$ | 252 | $B_{1g}$ | 257 | $B_{1g}$ |
| *15* | 263 | $B_{2g}$ | 273 | $B_{2g}$ | 249 | $B_{2g}$ | 250 | $B_{2g}$ |
| *16* | 234 | $A_g$ | 240 | $A_g$ | 225 | $A_g$ | 232 | $A_g$ |
| *17* | 195 | $B_{3u}$ | 203 | $B_{3u}$ | 212 | $A_u$ | 196 | $B_{1u}$ |
| *18* | 191 | $B_{1u}$ | 197 | $B_{1u}$ | 207 | $B_{1u}$ | 193 | $A_u$ |
| *19* | 188 | $A_u$ | 191 | $A_u$ | 204 | $B_{3u}$ | 193 | $B_{3u}$ |
| *20* | 181 | $B_{1g}$ | 186 | $B_{1g}$ | 184 | $B_{1g}$ | 179 | $B_{1g}$ |
| *21* | 168 | $A_u$ | 172 | $A_u$ | 174 | $B_{2u}$ | 173 | $B_{2u}$ |
| *22* | 166 | $B_{2u}$ | 172 | $B_{2u}$ | 168 | $B_{2u}$ | 168 | $A_u$ |
| *23* | 158 | $B_{2u}$ | 165 | $B_{2u}$ | 159 | $A_u$ | 160 | $B_{2u}$ |
| *24* | 146 | $B_{3u}$ | 148 | $B_{3u}$ | 139 | $B_{3g}$ | 141 | $B_{3u}$ |
| *25* | 139 | $B_{1u}$ | 148 | $B_{1u}$ | 134 | $B_{2g}$ | 137 | $B_{1u}$ |
| *26* | 136 | $B_{2g}$ | 137 | $B_{2g}$ | 132 | $B_{3u}$ | 135 | $B_{2g}$ |
| *27* | 127 | $B_{3g}$ | 136 | $B_{3g}$ | 123 | $B_{1u}$ | 119 | $B_{3g}$ |
| *28* | 102 | $A_u$ | 114 | $A_u$ | 116 | $A_u$ | 102 | $A_u$ |
| *29* | 97 | $B_{2u}$ | 105 | $B_{2u}$ | 88 | $A_g$ | 96 | $B_{3u}$ |
| *30* | 91 | $B_{3u}$ | 99 | $B_{3u}$ | 83 | $B_{2u}$ | 95 | $B_{2u}$ |
| *31* | 75 | $A_u$ | 90 | $A_u$ | 65 | $B_{1u}$ | 76 | $A_u$ |
| *32* | 69 | $A_g$ | 79 | $A_g$ | 64 | $B_{3u}$ | 68 | $A_g$ |
| *33* | 35 | $B_{1u}$ | 45 | $B_{1u}$ | 44 | $A_u$ | 31 | $B_{1u}$ |
| *34* | 0 | $B_{2u}$ | 0 | $B_{2u}$ | 15 | $B_{3u}$ | 1 | $B_{2u}$ |
| *35* | 0 | $B_{1u}$ | 1 | $B_{1u}$ | 24 | $B_{2u}$ | 2 | $B_{3u}$ |
| *36* | 1 | $B_{3u}$ | 1 | $B_{3u}$ | 29 | $B_{1u}$ | 3 | $B_{1u}$ |

**Table SI9.** Phonon frequencies for $CuF_2$ in *Pbca* system for all methods. The last three bands correspond to the translations.

| | DFT+U | | DFT+U+vdw | | SCAN | | HSE06 | |
|---|---|---|---|---|---|---|---|---|
| # | $cm^{-1}$ | Symm. | $cm^{-1}$ | Symm. | $cm^{-1}$ | Symm. | $cm^{-1}$ | Symm. |
| *1* | 523 | $A_u$ | 528 | $A_u$ | 527 | $A_u$ | 559 | $B_{1g}$ |
| *2* | 523 | $B_{1g}$ | 526 | $B_{1g}$ | 518 | $B_{3u}$ | 555 | $B_{2g}$ |
| *3* | 520 | $B_{2g}$ | 524 | $B_{2g}$ | 512 | $B_{1u}$ | 544 | $A_u$ |
| *4* | 513 | $B_{3u}$ | 518 | $B_{3u}$ | 490 | $B_{2u}$ | 536 | $B_{3u}$ |
| *5* | 504 | $B_{1u}$ | 511 | $B_{1u}$ | 476 | $B_{2g}$ | 528 | $B_{1u}$ |
| *6* | 473 | $B_{2u}$ | 476 | $B_{2u}$ | 475 | $B_{1g}$ | 507 | $B_{2u}$ |
| *7* | 427 | $B_{2u}$ | 432 | $B_{2u}$ | 410 | $B_{2u}$ | 437 | $B_{2u}$ |
| *8* | 386 | $B_{3g}$ | 395 | $A_u$ | 375 | $A_u$ | 393 | $A_u$ |
| *9* | 385 | $A_u$ | 388 | $B_{3g}$ | 374 | $B_{3g}$ | 391 | $B_{3u}$ |
| *10* | 366 | $B_{3u}$ | 372 | $B_{1u}$ | 372 | $B_{3u}$ | 391 | $B_{3g}$ |
| *11* | 366 | $B_{1u}$ | 371 | $B_{3u}$ | 369 | $B_{1u}$ | 388 | $B_{1u}$ |
| *12* | 360 | $B_{3g}$ | 369 | $B_{3g}$ | 348 | $A_g$ | 363 | $A_g$ |
| *13* | 349 | $A_g$ | 357 | $B_{2g}$ | 315 | $B_{3g}$ | 355 | $B_{3g}$ |
| *14* | 345 | $B_{2g}$ | 347 | $A_g$ | 314 | $B_{2g}$ | 337 | $B_{2g}$ |
| *15* | 312 | $B_{1g}$ | 318 | $B_{1g}$ | 293 | $B_{1g}$ | 315 | $B_{1g}$ |
| *16* | 274 | $B_{1g}$ | 283 | $B_{1g}$ | 243 | $B_{1g}$ | 263 | $B_{3u}$ |
| *17* | 261 | $B_{3u}$ | 267 | $B_{3u}$ | 241 | $B_{1u}$ | 260 | $B_{1g}$ |
| *18* | 255 | $B_{1u}$ | 261 | $B_{1u}$ | 241 | $B_{3u}$ | 259 | $B_{1u}$ |
| *19* | 248 | $A_u$ | 254 | $A_u$ | 235 | $A_g$ | 256 | $A_g$ |
| *20* | 238 | $A_g$ | 242 | $A_g$ | 232 | $A_u$ | 247 | $A_u$ |
| *21* | 236 | $B_{2u}$ | 241 | $B_{2u}$ | 225 | $B_{2u}$ | 235 | $B_{2u}$ |
| *22* | 229 | $A_u$ | 231 | $A_u$ | 216 | $A_u$ | 234 | $A_u$ |
| *23* | 217 | $B_{3g}$ | 231 | $B_{3g}$ | 200 | $B_{2u}$ | 211 | $B_{2u}$ |
| *24* | 198 | $B_{2u}$ | 203 | $B_{2u}$ | 187 | $B_{3g}$ | 198 | $B_{3g}$ |
| *25* | 167 | $B_{3u}$ | 180 | $A_g$ | 163 | $B_{3u}$ | 174 | $B_{3u}$ |
| *26* | 166 | $A_g$ | 172 | $B_{3u}$ | 158 | $B_{1u}$ | 165 | $B_{1u}$ |
| *27* | 157 | $A_u$ | 160 | $A_u$ | 143 | $B_{2g}$ | 156 | $B_{2g}$ |
| *28* | 154 | $B_{1u}$ | 160 | $B_{1u}$ | 142 | $A_g$ | 151 | $A_u$ |
| *29* | 153 | $B_{2g}$ | 157 | $B_{2g}$ | 140 | $A_u$ | 150 | $A_g$ |
| *30* | 126 | $B_{2u}$ | 141 | $B_{2u}$ | 127 | $B_{2u}$ | 125 | $B_{3u}$ |
| *31* | 113 | $B_{3u}$ | 120 | $B_{3u}$ | 123 | $B_{3u}$ | 124 | $B_{2u}$ |
| *32* | 91 | $A_u$ | 104 | $A_u$ | 100 | $A_u$ | 96 | $A_u$ |
| *33* | 54 | $B_{1u}$ | 71 | $B_{1u}$ | 49 | $B_{1u}$ | 56 | $B_{1u}$ |
| *34* | 1 | $B_{2u}$ | 1 | $B_{3u}$ | 1 | $B_{2u}$ | 1 | $B_{1u}$ |
| *35* | 1 | $B_{1u}$ | 1 | $B_{2u}$ | 3 | $B_{3u}$ | 2 | $B_{2u}$ |
| *36* | 2 | $B_{3u}$ | 2 | $B_{1u}$ | 4 | $B_{1u}$ | 2 | $B_{3u}$ |

**Table SI10.** Phonon frequencies for $AgF_2$ in $P2_1/c$ system for all methods. The last three bands correspond to the translations.

| | DFT+U | | DFT+U+vdW | | SCAN | | HSE06 | |
|---|---|---|---|---|---|---|---|---|
| # | $cm^{-1}$ | Symm. | $cm^{-1}$ | Symm. | $cm^{-1}$ | Symm. | $cm^{-1}$ | Symm. |
| 1 | 486 | $B_g$ | 493 | $B_g$ | 458 | $A_u$ | 487 | $B_g$ |
| 2 | 450 | $A_u$ | 453 | $B_u$ | 455 | $B_u$ | 463 | $A_u$ |
| 3 | 447 | $B_u$ | 452 | $A_u$ | 324 | $A_g$ | 460 | $B_u$ |
| 4 | 341 | $A_g$ | 355 | $A_g$ | 276 | $A_u$ | 348 | $A_g$ |
| 5 | 322 | $A_u$ | 333 | $A_u$ | 270 | $B_g$ | 316 | $A_u$ |
| 6 | 319 | $B_u$ | 326 | $B_u$ | 257 | $B_u$ | 316 | $B_u$ |
| 7 | 242 | $A_g$ | 241 | $B_g$ | 208 | $A_g$ | 239 | $A_g$ |
| 8 | 237 | $B_g$ | 240 | $A_g$ | 183 | $B_u$ | 233 | $B_g$ |
| 9 | 183 | $B_u$ | 183 | $B_u$ | 183 | $B_g$ | 188 | $B_u$ |
| 10 | 180 | $B_g$ | 180 | $A_u$ | 180 | $A_u$ | 178 | $B_g$ |
| 11 | 176 | $A_u$ | 175 | $B_g$ | 130 | $B_g$ | 178 | $A_u$ |
| 12 | 167 | $A_u$ | 158 | $A_u$ | 109 | $A_u$ | 165 | $A_u$ |
| 13 | 145 | $B_u$ | 149 | $B_u$ | 96 | $B_u$ | 140 | $B_u$ |
| 14 | 103 | $A_u$ | 108 | $A_u$ | 94 | $A_u$ | 103 | $A_u$ |
| 15 | 48 | $A_g$ | 54 | $A_g$ | 63 | $A_g$ | 59 | $A_g$ |
| 16 | 0 | $B_u$ | 0 | $B_u$ | 0 | $B_u$ | 0 | $B_u$ |
| 17 | 1 | $A_u$ | 0 | $A_u$ | 0 | $B_u$ | 0 | $A_u$ |
| 18 | 1 | $B_u$ | 1 | $B_u$ | 1 | $A_u$ | 2 | $B_u$ |

**Table SI11.** Phonon frequencies for $CuF_2$ in $P2_1/c$ system for all methods. The last three bands correspond to the translations.

| | DFT+U | | DFT+U+vdW | | SCAN | | HSE06 | |
|---|---|---|---|---|---|---|---|---|
| # | $cm^{-1}$ | Symm. | $cm^{-1}$ | Symm. | $cm^{-1}$ | Symm. | $cm^{-1}$ | Symm. |
| 1 | 551 | $B_g$ | 560 | $B_g$ | 503 | $A_u$ | 575 | $B_g$ |
| 2 | 493 | $A_u$ | 500 | $A_u$ | 495 | $B_u$ | 518 | $A_u$ |
| 3 | 482 | $B_u$ | 490 | $B_u$ | 492 | $B_g$ | 508 | $B_u$ |
| 4 | 380 | $A_g$ | 389 | $A_g$ | 371 | $A_g$ | 379 | $A_g$ |
| 5 | 363 | $A_u$ | 373 | $A_u$ | 362 | $A_u$ | 377 | $A_u$ |
| 6 | 351 | $B_u$ | 357 | $B_u$ | 349 | $B_u$ | 372 | $B_u$ |
| 7 | 292 | $B_g$ | 294 | $B_g$ | 266 | $B_g$ | 291 | $B_g$ |
| 8 | 258 | $A_g$ | 263 | $A_g$ | 237 | $A_g$ | 267 | $A_g$ |
| 9 | 247 | $A_u$ | 254 | $A_u$ | 229 | $B_u$ | 245 | $B_u$ |
| 10 | 246 | $B_u$ | 250 | $B_u$ | 215 | $A_u$ | 236 | $A_u$ |
| 11 | 232 | $B_g$ | 235 | $B_g$ | 208 | $A_u$ | 227 | $B_g$ |
| 12 | 220 | $A_u$ | 220 | $A_u$ | 198 | $B_g$ | 227 | $A_u$ |
| 13 | 188 | $B_u$ | 192 | $B_u$ | 167 | $B_u$ | 185 | $B_u$ |
| 14 | 140 | $A_u$ | 145 | $A_u$ | 127 | $A_u$ | 138 | $A_u$ |
| 15 | 84 | $A_g$ | 75 | $A_g$ | 59 | $A_g$ | 77 | $A_g$ |
| 16 | 1 | $B_u$ | 0 | $B_u$ | 1 | $A_u$ | 1 | $B_u$ |
| 17 | 2 | $A_u$ | 1 | $B_u$ | 1 | $B_u$ | 1 | $A_u$ |
| 18 | 2 | $B_u$ | 2 | $A_u$ | 1 | $B_u$ | 2 | $B_u$ |

## SI6. Electronic density of states for $AgF_2$/$CuF_2$ in $P2_1/c$ and *Pbca* forms from HSE06 method

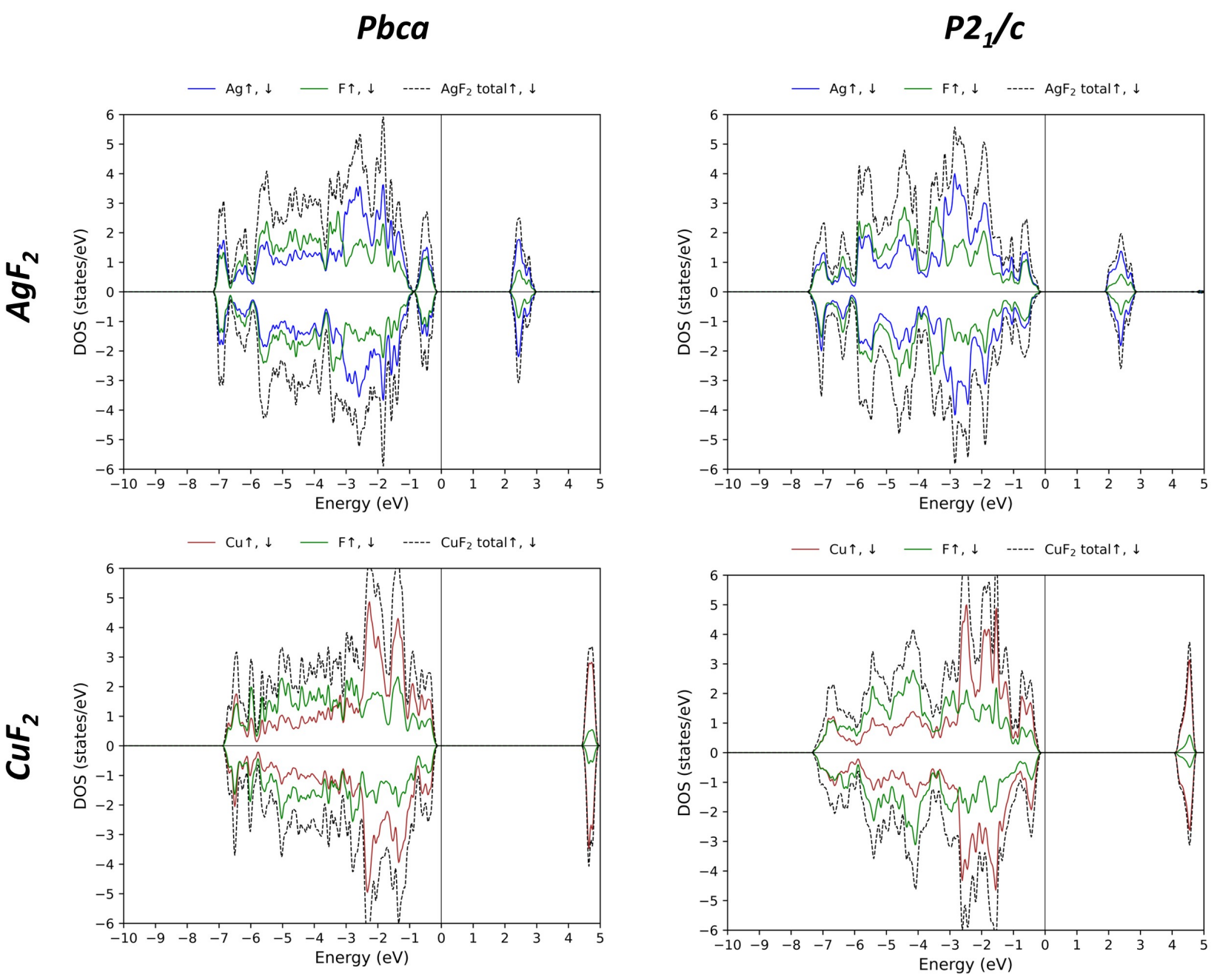

**Fig. SI1** Atom-resolved density of states for $AgF_2$ and $CuF_2$ in both crystal structures, within HSE06 method.

## SI7 Optimized structures from SCAN method

$AgF_2$_*Pbca*_DFT+U

```
   5.4325827684372898    0.0000000000000000    0.0000000000000000
   0.0000000000000000    5.7683784010416890    0.0000000000000000
   0.0000000000000000    0.0000000000000000    5.0079337523618577
  Ag   F
   4     8
Direct
 -0.0000000000000000 -0.0000000000000000 -0.0000000000000000
  0.5000000000000000 -0.0000000000000000  0.5000000000000000
 -0.0000000000000000  0.5000000000000000  0.5000000000000000
  0.5000000000000000  0.5000000000000000 -0.0000000000000000
  0.3081704282328906  0.8672726819052461  0.1810474953492010
  0.6918295717671094  0.1327272890947479  0.8189525046507991
  0.1918295717671094  0.1327272890947479  0.6810474953492009
  0.8081704282328906  0.8672726819052461  0.3189525046507990
  0.6918295717671094  0.3672727109052521  0.3189525046507990
  0.3081704282328906  0.6327273180947539  0.6810474953492009
  0.8081704282328906  0.6327273180947539  0.8189525046507991
  0.1918295717671094  0.3672727109052521  0.1810474953492010
```

$AgF_2$_*Pbca*_DFT+U+vdW

```
   5.4084161242709152    0.0000000000000000    0.0000000000000000
   0.0000000000000000    5.6604891530796744    0.0000000000000000
  -0.0000000000000000    0.0000000000000000    4.9925371834847416
  Ag   F
```

4 8
Direct
-0.0000000000000000 -0.0000000000000000 -0.0000000000000000
0.5000000000000000 -0.0000000000000000 0.5000000000000000
-0.0000000000000000 0.5000000000000000 0.5000000000000000
0.5000000000000000 0.5000000000000000 -0.0000000000000000
0.3112495951615513 0.8666956753121100 0.1776014607481890
0.6887504048384488 0.1333042956878842 0.8223985392518115
0.1887504048384489 0.1333042956878842 0.6776014607481885
0.8112495951615512 0.8666956753121100 0.3223985392518112
0.6887504048384488 0.3666957043121157 0.3223985392518112
0.3112495951615513 0.6333043246878900 0.6776014607481885
0.8112495951615512 0.6333043246878900 0.8223985392518115
0.1887504048384489 0.3666957043121157 0.1776014607481890

$AgF_2$_*Pbca*_SCAN
5.5825253940946036 0.0000000000000000 0.0000000000000000
0.0000000000000000 5.7238028965063732 0.0000000000000000
0.0000000000000000 0.0000000000000000 5.1506410134567648
Ag F
4 8
Direct
-0.0000000000000000 -0.0000000000000000 0.0000000000000000
0.5000000000000000 -0.0000000000000000 0.5000000000000000
-0.0000000000000000 0.5000000000000000 0.5000000000000000
0.5000000000000000 0.5000000000000000 0.0000000000000000
0.3018553765971723 0.8676807184337415 0.1894365447265209
0.6981446234028278 0.1323192525662527 0.8105634552734791
0.1981446234028275 0.1323192525662527 0.6894365447265209
0.8018553765971722 0.8676807184337415 0.3105634552734791
0.6981446234028278 0.3676807474337472 0.3105634552734791
0.3018553765971723 0.6323192815662585 0.6894365447265209
0.8018553765971722 0.6323192815662585 0.8105634552734791
0.1981446234028275 0.3676807474337472 0.1894365447265210

$AgF_2$_*Pbca*_HSE06
5.5385447024405279 0.0000000000000000 -0.0000000000000000
0.0000000000000000 5.8425028417607319 0.0000000000000000
-0.0000000000000000 0.0000000000000000 5.0823121733048504
Ag F
4 8
Direct
0.0000000000000000 -0.0000000000000000 -0.0000000000000000
0.5000000000000000 -0.0000000000000000 0.5000000000000000
0.0000000000000000 0.5000000000000000 0.5000000000000000
0.5000000000000000 0.5000000000000000 -0.0000000000000000
0.3035307753632584 0.8727315986283707 0.1861317494483685
0.6964692246367415 0.1272683723716233 0.8138682505516314
0.1964692246367415 0.1272683723716233 0.6861317494483686
0.8035307753632585 0.8727315986283707 0.3138682505516315
0.6964692246367415 0.3727316276283767 0.3138682505516315
0.3035307753632584 0.6272684013716293 0.6861317494483686
0.8035307753632585 0.6272684013716293 0.8138682505516314

0.1964692246367415 0.3727316276283767 0.1861317494483685

$AgF_2$_$P2_1/c$_DFT+U
3.6384468137734780 0.0000000000000000 -0.0041520798417513
0.0000000000000000 4.7768034691478833 0.0000000000000000
-3.0257122287743710 0.0000000000000000 4.8372583721165396
Ag F
2 4
Direct
0.0000000000000000 -0.0000000000000000 0.0000000000000000
0.0000000000000000 0.5000000000000000 0.5000000000000000
0.2580612161134505 0.3032513596006569 0.2979143887080884
0.7419387838865490 0.6967486113993373 0.7020856402919100
0.7419387838865490 0.8032513886006627 0.2020856112919114
0.2580612161134505 0.1967486403993431 0.7979143597080900

$AgF_2$_$P2_1/c$_ DFT+U+vdW
3.6578695608297194 0.0000000000000000 -0.0449210475348461
0.0000000000000000 4.7214582353131362 0.0000000000000000
-3.0965754060751922 0.0000000000000000 4.7800051707746958
Ag F
2 4
Direct
0.0000000000000000 -0.0000000000000000 0.0000000000000000
0.0000000000000000 0.5000000000000000 0.5000000000000000
0.2601460237587488 0.3070580012521441 0.2989172378806293
0.7398539762412513 0.6929419697478429 0.7010827911193696
0.7398539762412513 0.8070580302521571 0.2010827621193707
0.2601460237587488 0.1929419987478489 0.7989172088806304

$AgF_2$_$P2_1/c$_SCAN
3.7415840779467833 0.0000000000000000 -0.0918547643182602
0.0000000000000000 4.9088817666584550 0.0000000000000000
-3.1729438680646593 0.0000000000000000 4.8882392413909601
Ag F
2 4
Direct
-0.0000000000000000 -0.0000000000000000 -0.0000000000000000
-0.0000000000000000 0.5000000000000000 0.5000000000000000
0.2460597954672760 0.3072026777366762 0.2961816016485471
0.7539402045327240 0.6927972932633180 0.7038184273514519
0.7539402045327240 0.8072027067366820 0.2038183983514530
0.2460597954672760 0.1927973222633238 0.7961815726485481

$AgF_2$_$P2_1/c$_HSE06
3.7170509955380622 0.0000000000000000 0.0047660820843200
0.0000000000000000 4.8239178583612023 0.0000000000000000
-3.0792251184526580 0.0000000000000000 4.8870756396050909
Ag F
2 4
Direct

0.0000000000000000 -0.0000000000000000 -0.0000000000000000
0.0000000000000000 0.5000000000000000 0.5000000000000000
0.2505233569641341 0.3029229041044571 0.2967999047589231
0.7494766430358657 0.6970770668955373 0.7032001242410757
0.7494766430358657 0.8029229331044627 0.2032000952410770
0.2505233569641341 0.1970770958955429 0.7967998757589243

$CuF_2$_*Pbca*_DFT+U
4.9711335115512245 0.0000000000000000 -0.0000000000000000
0.0000000000000000 5.3097794773679450 0.0000000000000000
0.0000000000000000 0.0000000000000000 4.6679476639154105
Cu F
4 8
Direct
-0.0000000000000000 -0.0000000000000000 -0.0000000000000000
0.5000000000000000 -0.0000000000000000 0.5000000000000000
-0.0000000000000000 0.5000000000000000 0.5000000000000000
0.5000000000000000 0.5000000000000000 -0.0000000000000000
0.3183095129465213 0.8732020347780225 0.1717660482612136
0.6816904870534715 0.1267979362219716 0.8282339517387864
0.1816904870534716 0.1267979362219716 0.6717660482612136
0.8183095129465285 0.8732020347780225 0.3282339517387864
0.6816904870534715 0.3732020637780284 0.3282339517387864
0.3183095129465213 0.6267979652219775 0.6717660482612136
0.8183095129465285 0.6267979652219775 0.8282339517387864
0.1816904870534716 0.3732020637780284 0.1717660482612136

$CuF_2$_*Pbca*_DFT+U+vdW
4.9511571524161839 0.0000000000000000 -0.0000000000000000
0.0000000000000000 5.2592024115207971 0.0000000000000000
-0.0000000000000000 0.0000000000000000 4.6562936912203581
Cu F
4 8
Direct
-0.0000000000000000 0.0000000000000000 0.0000000000000000
0.5000000000000000 0.0000000000000000 0.5000000000000000
-0.0000000000000000 0.5000000000000000 0.5000000000000000
0.5000000000000000 0.5000000000000000 0.0000000000000000
0.3206705495773565 0.8731719024696725 0.1695527903438375
0.6793294504226364 0.1268280685303217 0.8304472096561626
0.1793294504226365 0.1268280685303217 0.6695527903438374
0.8206705495773636 0.8731719024696725 0.3304472096561625
0.6793294504226364 0.3731719314696783 0.3304472096561625
0.3206705495773565 0.6268280975303275 0.6695527903438374
0.8206705495773636 0.6268280975303275 0.8304472096561626
0.1793294504226365 0.3731719314696783 0.1695527903438375

$CuF_2$_*Pbca*_SCAN
5.0470466179036055 0.0000000000000000 -0.0000000000000000
0.0000000000000000 5.4448582294964751 0.0000000000000000
0.0000000000000000 0.0000000000000000 4.6802399712243048

```
  Cu   F
    4    8
Direct
  0.0000000000000000 -0.0000000000000000  0.0000000000000000
  0.5000000000000000 -0.0000000000000000  0.5000000000000000
  0.0000000000000000  0.5000000000000000  0.5000000000000000
  0.5000000000000000  0.5000000000000000  0.0000000000000000
  0.3152406344036440  0.8784445413341972  0.1731180155087471
  0.6847593655963489  0.1215554296657968  0.8268819844912528
  0.1847593655963489  0.1215554296657968  0.6731180155087472
  0.8152406344036511  0.8784445413341972  0.3268819844912529
  0.6847593655963489  0.3784445703342033  0.3268819844912529
  0.3152406344036440  0.6215554586658028  0.6731180155087472
  0.8152406344036511  0.6215554586658028  0.8268819844912527
  0.1847593655963489  0.3784445703342033  0.1731180155087471
```

$CuF_2$_*Pbca*_HSE06

```
     5.0375200457129949    0.0000000000000000    0.0000000000000000
     0.0000000000000000    5.4703986051368156    0.0000000000000000
    -0.0000000000000000    0.0000000000000000    4.6703820614650837
  Cu   F
    4    8
Direct
 -0.0000000000000000  0.0000000000000000  0.0000000000000000
  0.5000000000000000  0.0000000000000000  0.5000000000000000
  0.0000000000000000  0.5000000000000000  0.5000000000000000
  0.5000000000000000  0.5000000000000000  0.0000000000000000
  0.3129443861067328  0.8788919183021422  0.1760803685437572
  0.6870556138932602  0.1211080526978519  0.8239196314562428
  0.1870556138932601  0.1211080526978519  0.6760803685437572
  0.8129443861067398  0.8788919183021422  0.3239196314562429
  0.6870556138932602  0.3788919473021410  0.3239196314562429
  0.3129443861067328  0.6211080816978578  0.6760803685437572
  0.8129443861067398  0.6211080816978578  0.8239196314562428
  0.1870556138932601  0.3788919473021410  0.1760803685437572
```

$CuF_2$_$P2_1/c$_ DFT+U

```
     3.1918032843702164    0.0000000000000000   -0.0282795905905217
     0.0000000000000000    4.5049849400793818    0.0000000000000000
    -2.7117339860874119    0.0000000000000000    4.5458609099431406
  Cu   F
    2    4
Direct
  0.0000000000000000  0.0000000000000000  0.0000000000000000
  0.0000000000000000  0.5000000000000000  0.5000000000000000
  0.2633370751632447  0.2980220513899945  0.2958083607980840
  0.7366629248367553  0.7019779196100067  0.7041916682019148
  0.7366629248367553  0.7980220803899933  0.2041916392019160
  0.2633370751632447  0.2019779486100055  0.7958083317980852
```

$CuF_2$_$P2_1/c$_ DFT+U+vdW

```
     3.1843616963673211    0.0000000000000000   -0.0401347544237979
```

0.0000000000000000 4.4787430856724946 0.0000000000000000
-2.7224144821479328 0.0000000000000000 4.5312785681204444
Cu F
2 4
Direct
0.0000000000000000 -0.0000000000000000 0.0000000000000000
0.0000000000000000 0.5000000000000000 0.5000000000000000
0.2650117692506215 0.2983719618346402 0.2960198521911562
0.7349882307493785 0.7016280091653607 0.7039801768088426
0.7349882307493785 0.7983719908346393 0.2039801478088439
0.2650117692506215 0.2016280381653597 0.7960198231911574

$CuF_2$_P$2_1$/*c*_SCAN
3.3024484670833067 0.0000000000000000 0.0040032864862328
0.0000000000000000 4.5224078706716684 0.0000000000000000
-2.7558793437961513 0.0000000000000000 4.5385905413277614
Cu F
2 4
Direct
0.0000000000000000 0.0000000000000000 -0.0000000000000000
0.0000000000000000 0.5000000000000000 0.5000000000000000
0.2544117004725763 0.2984925719792934 0.2940252647869827
0.7455882995274238 0.7015073990207077 0.7059747642130163
0.7455882995274238 0.7984926009792923 0.2059747352130173
0.2544117004725763 0.2015074280207066 0.7940252357869837

$CuF_2$_P$2_1$/*c*_ HSE06
3.2956089918571907 0.0000000000000000 0.0114348804553170
0.0000000000000000 4.5164592600711346 0.0000000000000000
-2.7434020870818063 0.0000000000000000 4.5343499734704871
Cu F
2 4
Direct
-0.0000000000000000 0.0000000000000000 -0.0000000000000000
-0.0000000000000000 0.5000000000000000 0.5000000000000000
0.2530356795178663 0.2976717097460082 0.2947272241957566
0.7469643204821337 0.7023282612539932 0.7052728048042424
0.7469643204821337 0.7976717387460068 0.2052727758042435
0.2530356795178663 0.2023282902539919 0.7947271951957576

## SI8. Bibliography